%% file: main.tex
\documentclass[journal=tches,final]{iacrtrans}
\input{settings.tches}

\usepackage{booktabs}   %% For formal tables:
\usepackage{subcaption} %% For complex figures with subfigures/subcaptions
\DeclareMathOperator*{\argmin}{arg\,min}
\DeclareMathOperator*{\Total}{\sum}
\newcommand{\D}{\mathrm{d}}

\input{preamble}

\author{Ferhat Erata \and Chuanqi Xu \and Ruzica Piskac \and Jakub Szefer}
\institute{
  Yale University, New Haven, CT, US \\
  \email[ferhat.erata@yale.edu,chuanqi.xu@yale.edu,ruzica.piskac@yale.edu,jakub.szefer@yale.edu]{{firstname.lastname}@yale.edu}
}
\title[Quantum Circuit Reconstruction from Power Side-Channel Attacks]{Quantum Circuit Reconstruction from Power Side-Channel Attacks on Quantum Computer Controllers}
\begin{document}

\maketitle

%%%% 5. KEYWORDS %%%%

%%%% 6. ABSTRACT %%%%
\input{sections/abstract}

%%%% 7. PAPER CONTENT %%%%
\section{Introduction} \label{sec::introduction}
\input{sections/introduction.tex}

\section{Background} \label{sec::background}
\input{sections/background_quantum.tex}

\section{\revision{Attack Scenario and Threat Model}} 
\label{sec::threat_model}
\input{sections/threat_model.tex}

\section{Formalization of \revision{Circuit} Reconstruction} 
\label{sec::formalization}

\input{sections/formalization.tex}

\section{\revision{Power Side-Channel Attacks}} 
\label{sec::method}

\input{sections/method.tex}

\section{Evaluation Setup} 
\label{sec::evaluation}
\input{sections/evaluation.tex}

\section{\revision{Discussion and Future Work}}
\label{sec::future_work}
\input{sections/future_work.tex}

\section{Related Work} \label{sec::related_work}
\input{sections/related_work.tex}

\section{Conclusion} \label{sec::conclusion}
\input{sections/conclusion}

\section*{Acknowledgments}

This work was supported in part by National Science Foundation grants \nsf{2312754} and~\nsf{2245344}.

%%%% 8. BILBIOGRAPHY %%%%
\bibliographystyle{alpha}
\bibliography{bibliography}
%%%% NOTES
% - Download abbrev3.bib and crypto.bib from https://cryptobib.di.ens.fr/
% - Use bilbio.bib for additional references not in the cryptobib database.
%   If possible, take them from DBLP.

\end{document}

%% file: settings.tches.tex
\setfirstpage{735}
\setlastpage{768}
\setvolume{2024}
\setnumber{2}

\setISSN{2569-2925}
\makeatletter
\setDOI{10.46586/tches.v\IACR@vol.i\IACR@no.\IACR@fp-\IACR@lp}
\makeatother

\setkeys{IACR}{journal=tches}

%% file: preamble.tex
\usepackage{enumitem}
\usepackage{caption}

\definecolor{darkgreen}{rgb}{0.0, 0.5, 0.0}

\usepackage{amsmath}
\usepackage{soul}
\usepackage[braket, qm]{qcircuit}
\usepackage{physics}
\usepackage{graphicx}
\usepackage{enumitem}
\usepackage{multirow}
\usepackage{hyperref}

\usepackage{xcolor}
\definecolor{darkgreen}{rgb}{0.0, 0.5, 0.0}
\long\def\revision#1{{\texorpdfstring{}{}#1}}

\usepackage{balance}

\usepackage{algorithm}
\usepackage{algpseudocode}

\usepackage{nicematrix}
\usepackage{wrapfig}

\usepackage{lipsum}

\newcommand{\nsf}[1]{\href{https://www.nsf.gov/awardsearch/showAward?AWD_ID=#1}{#1}}

%% file: sections/abstract.tex
\begin{abstract}

The interest in quantum computing has grown rapidly in recent years, and with it grows the importance of securing quantum circuits. A novel type of threat to quantum circuits that dedicated attackers could launch are power trace attacks. To address this threat, this paper presents first formalization and demonstration of using power traces to unlock and steal quantum circuit secrets. With access to power traces, attackers can recover information about the control pulses sent to quantum computers. From the control pulses, the gate level description of the circuits, and eventually the secret algorithms can be reverse engineered. This work demonstrates how and what information could be recovered. This work uses algebraic reconstruction from power traces to realize two new types of single trace attacks: per-channel and total power attacks. The former attack relies on per-channel measurements to perform a brute-force attack to reconstruct the quantum circuits. The latter attack performs a single-trace attack using Mixed-Integer Linear Programming optimization. Through the use of algebraic reconstruction, this work demonstrates that quantum circuit secrets can be stolen with high accuracy. Evaluation on 32 real benchmark quantum circuits shows that our technique is highly effective at reconstructing quantum circuits. The findings not only show the veracity of the potential attacks, but also the need to develop new means to protect quantum circuits from power trace attacks. Throughout this work real control pulse information from real quantum computers is used to demonstrate potential attacks based on simulation of collection of power traces.

\end{abstract}

%% file: sections/introduction.tex
\begin{figure}[t]
  \centering
  \includegraphics[width=\columnwidth]{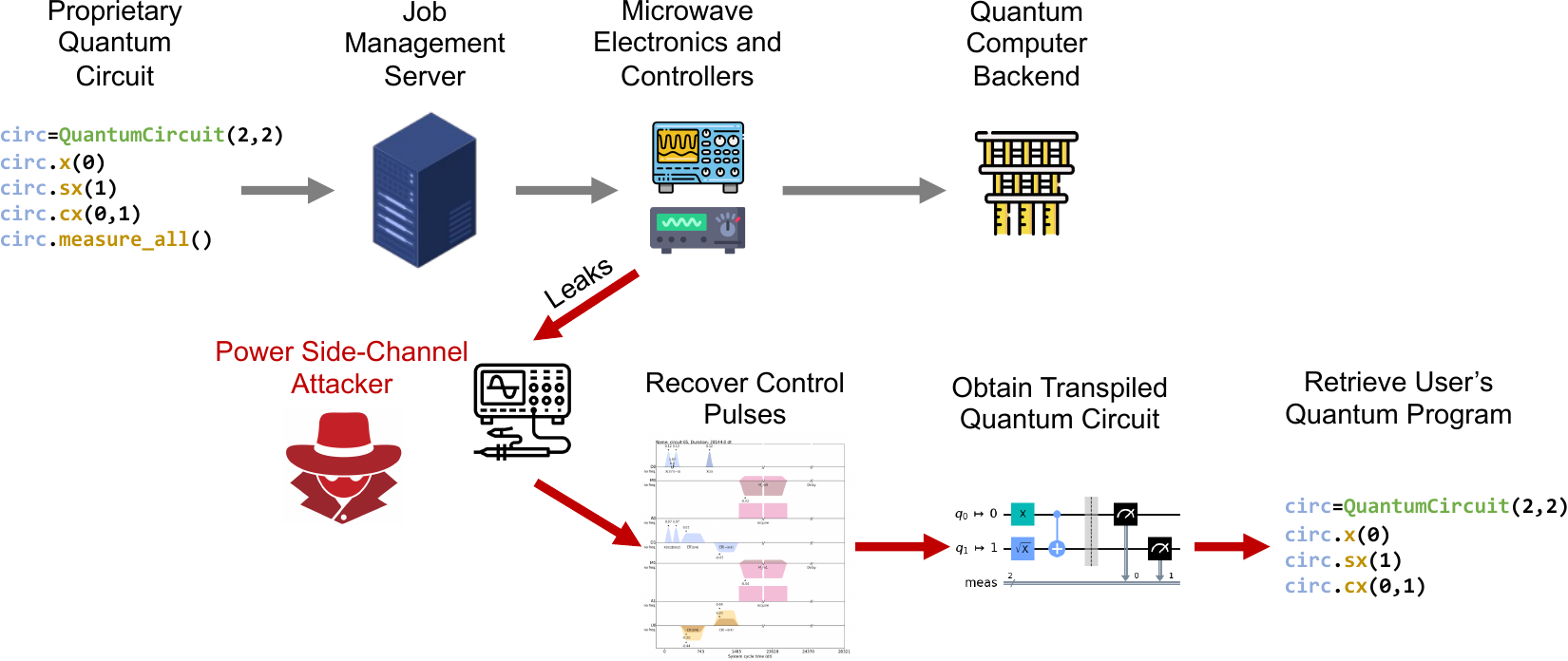}
  \caption{{Typical operation of a cloud-based quantum computer. Red arrows highlight potential power trace~threats.}}
  \label{fig_side_channel_threats}
\end{figure}

The interest in quantum computing is growing rapidly and already a large numbers of quantum computers are easily accessible over the internet to researchers and everyday users. Due to the expensive nature of the quantum computing equipment, these computers are currently available as cloud-based systems. For example, IBM Quantum~\cite{ibm_quantum}, Amazon Braket~\cite{braket}, and Microsoft Azure~\cite{azure} already provide access to various types of Noisy Intermediate-Scale Quantum (NISQ) devices from different vendors. Remote access makes it easy for different users and companies to run algorithms on real quantum computers without the need to purchase or maintain them. On the other hand, the users have no control over the physical space where the quantum computers are. While the cloud providers may not be themselves malicious, the threat of malicious insiders within data centers or cloud computing facilities is well-known in classical security~\cite{szefer2014cyber}. These malicious insiders may have physical access to the equipment of quantum computers. With access to the quantum computers and the microwave controllers, malicious insiders could leverage physically collected information to steal or leak the quantum circuit~secrets.

For classical computers, side-channel attacks of different types are a well-known threat~\cite{szefer2018principles}. Two widely studied and analyzed side-channels are timing- and power-based channels. There are also thermal, EM, acoustic, and a variety of other categories of side-channels.
In timing side-channel attacks the attacker is trying to learn some secret properties about the circuit by measuring the execution times.
Timing attacks are powerful enough in classical computers to break the implementation of standard cryptographic primitives, such as DSA, RSA or Diffie-Hellman~\cite{DBLP:conf/crypto/Kocher96}. Timing side-channels are easier to exploit as they only require doing timing measurement of the victim.
Power side-channel attacks are more convoluted attacks, where the attacker tries to establish
a correlation between the power consumption and the operations and data that the circuit executes.
Kocher et al.~\cite{DBLP:conf/crypto/KocherJJ99} showed how to reconstruct the encryption keys in the Data Encryption Standard (DES) using a power consumption analysis. Power attacks require physical access to monitor the execution of the target computer.

Power side channels are well studied for classical computers~\cite{erata2023towards, prouff2013masking, synthesis2021icse, synthesis2021date, synthesis2020tornado, synthesis2019tcad, synthesis2019fse, synthesis2018taco, synthesis2017post, synthesis2015tc, synthesis2014cav, synthesis2012ches}. There are even platforms~\cite{chipwhisperer}
for analyzing power side channels. However, understanding power side channels for quantum computers
has not been explored yet, which this work aims to address. We show how a malicious
attacker can reconstruct a secret quantum circuit that is being executed,
by simply measuring the power consumption.

There is very limited research on understanding power side-channel attacks on quantum computers~\cite{xu2023exploration}. One insight about quantum computers is that if the attacker is to perform physical measurements on the qubits during the  computation, these measurements would interact with the qubits and destroy their state. However, we observe that each quantum computer is controlled by external hardware such as microwave electronics and controllers. Quantum computers, such as superconducting qubit machines from IBM, Rigetti, or others, use microwave pulses to execute gate operations on qubits. The control pulses are fully classical and could be spied on -- which is the target of this work.

This paper shows how by measuring the power consumption of the controller devices sending microwave pulses to a quantum computer, we can recover a potentially secret quantum circuit that the quantum computer executed. We show that anybody with access to power traces of the control pulse generation devices can capture and recover the control information. While this work explores power-based side-channels, the same or similar ideas could apply to electromagnetic (EM) or other types of physical side-channels. This is left as orthogonal work.

\subsection{Power {Side-Channel} Threats to Quantum Computers}

Figure~\ref{fig_side_channel_threats} shows the operation of today's cloud-based quantum computers. Remote users submit jobs to the cloud provider, where the job management or similar server dispatches the jobs to particular quantum computers, also called backends on IBM Quantum. Typically the digital instructions are sent to controller logic, such as microwave electronics, which generate the actual analog control signals sent to the quantum computer.

We assume that the classical computer components, e.g., the job management server, are protected from side-channels. Meanwhile, controller electronics of quantum computers, such as arbitrary waveform generators (AWGs) have not been analyzed for potential side-channels before this work. Consequently, we focus on and demonstrate potential new, power {side-channel} attacks that could be used to extract information about users' quantum circuits {(quantum gates and qubits)}. Rather than targeting the superconducting qubits themselves (which are isolated in a cryogenic refrigerator), we focus on the controller electronics shown in the middle of Figure~\ref{fig_side_channel_threats}.

The vulnerabilities in quantum computer controllers encompass more than just gate recovery. An attacker might discern the number of qubits used in a quantum program, a significant concern for algorithms like Quantum Approximate Optimization Algorithm (QAOA)~\cite{choi2019tutorial}. Variational quantum algorithms~\cite{cerezo2021variational} are notably sensitive to qubit count and circuit depth. Extracting such hyperparameters can reveal crucial information about the algorithm. Moreover, given the current state of quantum computing where inputs are hard-coded, an attacker could potentially extract sensitive input data from the~circuit.

Our primary focus is to protect the intellectual property embedded in quantum programs. Quantum circuits encapsulate both the algorithm and its inputs. An attacker capable of recovering parts of the quantum circuit can misappropriate this intellectual property or sensitive data.
This concern amplifies when quantum programs run on external quantum computers. A recent workshop by The National Quantum Coordination Office underscored this issue, emphasizing the role of formal methods in enhancing quantum computing security~\cite{nqco2022}. Our research aligns with this perspective, pioneering the use of formal methods for quantum circuit recovery.

\subsection{Lessons from Historical Technological Threats}

The significance of vulnerabilities in quantum computer controllers cannot be overstated, considering the ongoing evolution of technological advancements and their inherent risks. For perspective, speculative execution attacks in classical computers were not recognized as threats until 2018~\cite{lipp2020meltdown,kocher2020spectre}, despite having been operational since the commercialization of the technology by IBM and Intel in the 1990s. This delay between innovation and the identification of vulnerabilities underscores the necessity for proactive security measures.

In the context of quantum computers, addressing vulnerabilities in their infancy is crucial, even as the field continues to develop. Our research delves into these potential risks, emphasizing the importance of safeguarding quantum computing systems.

Drawing from past lessons, unchecked technological vulnerabilities can culminate in substantial security breaches. As cache attacks emerged long after caches were invented, a similar oversight with quantum computers could be costly. It's imperative to address these challenges proactively, ensuring quantum computing's advancement aligns with rigorous security protocols.

\subsection{Contributions}
Compared to the work by Xu et al.~\cite{xu2023exploration}, our work provides a formalization of the power side channel attack. Additionally, we also present a novel algebraic reconstruction method for recovery of quantum circuits. There are two reconstruction methods that we introduce; those methods depend on the attackers' abilities. In the per-channel method we assume that the attacker collects power traces from individual qubit channels. Meanwhile, in our total power single trace method we assume the attacker can only measure the total power trace of all~channels.

We have empirically evaluated our approach for 32 benchmark quantum circuits. The evaluation shows that our technique is highly effective at reconstructing quantum circuits.
In summary, the paper contributes the following:

\begin{itemize}
  \item The first formalization of power side channel attacks on complete reconstruction of quantum circuits from power traces, which is given in Section~\ref{sec::formalization}.
  \item Demonstration of {circuit reconstruction} using our new per-channel single trace attack: this attack relies on {single-shot} per-channel measurements to perform a brute-force attack to reconstruct the quantum circuits, which is given in Section~\ref{sec:per-channel-attacker}.
  \item Demonstration of {circuit reconstruction} using our new {single-shot total power side-channel} attack: this attack relies on a single power side-channel measurement and performs attack using Linear Mixed Integer Real Arithmetic (LIRA) solving and Mixed-Integer Linear Programming (MILP) optimization, which is given in Section~\ref{sec:total-power-attacker}.
  \item Details of the evaluation of the attacks on 32 real quantum circuits in the QASMBench\footnote{\url{https://github.com/pnnl/QASMBench/}} benchmark suite~\cite{qasmbench}, using control pulse information from real IBM quantum computers, which is given in Section~\ref{sec::evaluation}.
\end{itemize}

%% file: sections/background_quantum.tex
This section provides background on quantum computers and typical quantum computer~workflow.

\begin{figure*}[t]
  \centering
  \includegraphics[width=\linewidth]{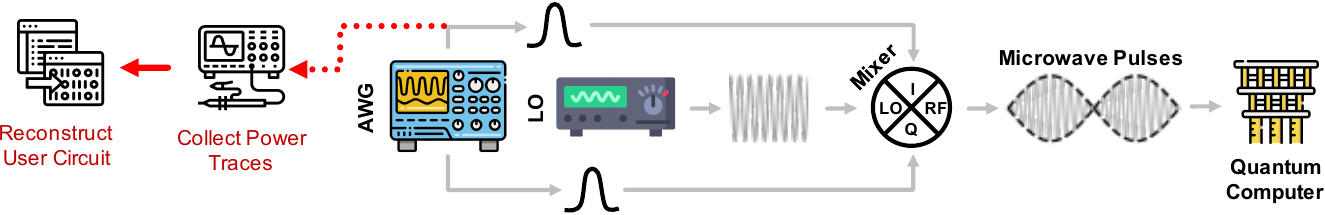}
  \caption[Schematic of a typical qubit drive setup]{Schematic of a typical qubit drive setup. The local oscillator (LO) generates a low phase-noise microwave carrier signal, and then the wave is modulated in the IQ mixer by I and Q components generated by the arbitrary wave generator (AWG). The pulse is then sent to drive the qubits in the quantum computer. The red line shows the process to collect power traces, which can be exploited by attackers to retrieve information.}
  \label{fig_hardware_sketch}
\end{figure*}

\subsection{Qubits and Quantum States}

The most essential component of quantum computing is the quantum bit, or qubit for short. It is theoretically comparable to the bit used in current classical computing. Like a bit, a qubit has two basis states, which are represented by the bra-ket notation as $\ket 0$ and $\ket 1$. Whereas a classical bit can only be either 0 or 1, a qubit can be any linear combination of $\ket 0$ and $\ket 1$ with norm 1. With this notation, a qubit $\ket \psi$ is typically represented as:
  $\ket \psi = \alpha \ket 0 + \beta \ket 1$, where  $\alpha$  and $\beta$ are complex numbers satisfying $|\alpha|^2 + |\beta|^2 = 1$.
Qubits are frequently represented using vectors. For example, $\ket 0 = [1, 0]^T$ and $\ket 1 = [0, 1]^T$ are two-dimensional vectors that can be used to represent the basis states of a single qubit. The above state $\ket \psi$ can therefore be expressed as $\ket \psi = \alpha \ket 0 + \beta \ket 1 = [\alpha, \beta]^T$. There are representations that are equivalent for multi-qubit states. For instance, the two-qubit states' space is made up of the four basis states $\ket{00}$, $\ket{01}$, $\ket{10}$, and $\ket{11}$. In the space of $n$-qubit states, there are $2^n$ basis states that range from $\ket{0\dots 0}$ to $\ket{1\dots 1}$, and a $n$-qubit state $\ket \phi$ can be expressed as follows:
\begin{equation*}
  \ket \phi = \sum_{i = 0}^{2^n - 1} a_i \ket i \text{ where } \sum_{i = 0}^{2^n - 1}|a_i|^2 = 1.
\end{equation*}

\subsection{Quantum Gates}

The fundamental quantum operations at the logic-level are quantum gates, which are comparable to classical computing. Quantum algorithms are made up of a series of quantum gates that can convert input qubits into different quantum states. Quantum gates are unitary operations that modify the input qubits, i.e., for a quantum gate $U$ that is applied to a quantum state $\ket \psi$, the quantum state is evolved to $\ket \psi \rightarrow U \ket \psi$, and $U U^\dagger = U^\dagger U = I$. With the vector-matrix representation, $2^n \times 2^n$ matrices can be utilized to express $n$-qubit quantum gates.

One classical example is the gate that is analogous to the NOT gate in classical computing, the Pauli-{\tt X} gate, that exchanges the components of $\ket 0$ and $\ket 1$. Another significant example is the two-qubit CNOT gate, also known as the {\tt CX} gate, which, if the control qubit is in the state $\ket 1$, applies a Pauli-$X$ gate to the target qubit; otherwise, nothing happens. There are some more matrices of quantum gates along with their matrix representations. One thing to keep in mind is that our qubit order is consistent with that in Qiskit~\cite{Qiskit}, where the leftmost qubit is the most important and the rightmost qubit is the least important. As a result, if a different qubit order is used in other studies, the {\tt CX} gate may have a different matrix representation. Below we show several matrices of quantum gates:

{\small

\begin{equation*}
  {\tt I}=\begin{bmatrix}
    1 & 0 \\
    0 & 1
  \end{bmatrix},\
  {\tt X}=\begin{bmatrix}
    0 & 1 \\
    1 & 0
  \end{bmatrix},\
  {\tt CX} = \begin{bmatrix}
    1 & 0 & 0 & 0 \\
    0 & 0 & 0 & 1 \\
    0 & 0 & 1 & 0 \\
    0 & 1 & 0 & 0
  \end{bmatrix},\
  {\tt RZ(\theta)}=\begin{bmatrix}
    e^{-i\frac{\theta}{2}} & 0                     \\
    0                      & e^{i\frac{\theta}{2}}
  \end{bmatrix},\
  {\tt SX}=\frac{1}{2}\begin{bmatrix}
    1+i & 1-i \\
    1-i & 1+i
  \end{bmatrix}
\end{equation*}

}

A small number of quantum gates can be used to approximate any unitary quantum gate within a small error, as shown in the study~\cite{doi:10.1098/rspa.1995.0065}. 
One of the crucial configurations of quantum computers is the basis gates, also known as native gates. Different manufacturers or even various versions of quantum computers from same manufacturer may have different native gates. Choice of supporting different types of basis gates is a trade-off between numerous attributes like error rate and efficiency. Our experiments in this study were done on IBM Quantum, and typically, the basis gates are provided by IBM quantum computers are: {\tt I}, {\tt RZ}, {\tt SX}, {\tt X}, and {\tt CX}. Prior to being executed on physical quantum computing hardware, quantum gates such as the commonly utilized Hadamard gate need to be broken down into these basis gates.

\subsection{Control Pulses}
\label{sec::superconducting_quantum_computer_controls}

\begin{figure}[t]
  \centering
  \begin{subfigure}[t]{0.48\textwidth}
    \includegraphics[width=\columnwidth]{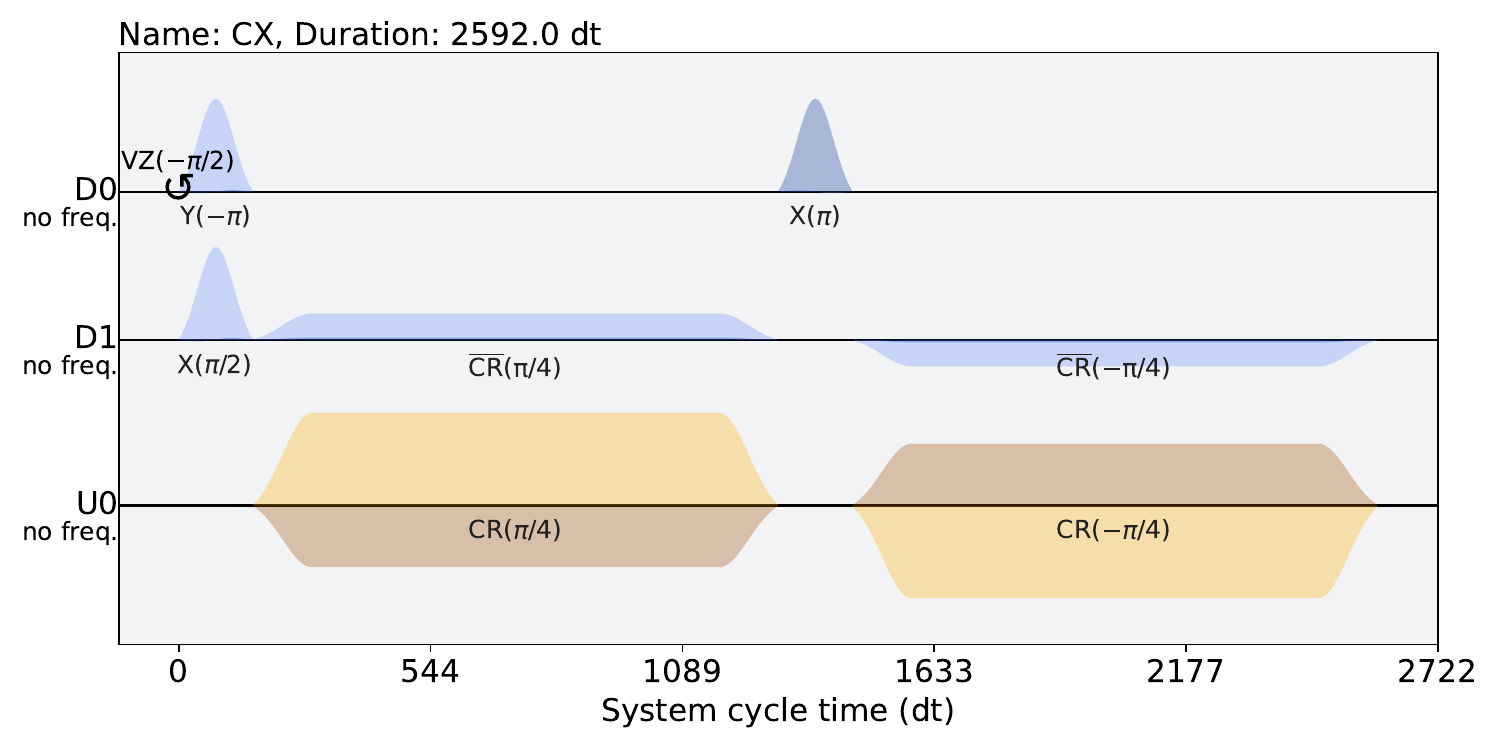}
    \caption{\footnotesize {\tt CX} pulse}
  \end{subfigure}
  \begin{subfigure}[t]{0.24\textwidth}
    \includegraphics[width=\columnwidth]{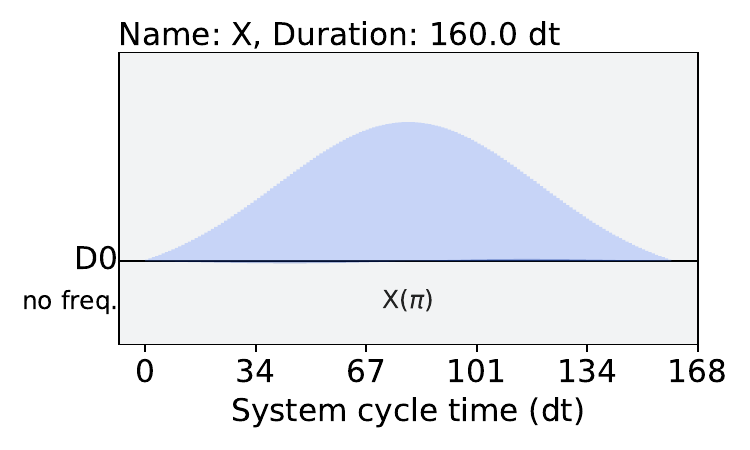}
    \caption{\footnotesize {\tt X} pulse}
  \end{subfigure}
  \begin{subfigure}[t]{0.24\textwidth}
    \includegraphics[width=\columnwidth]{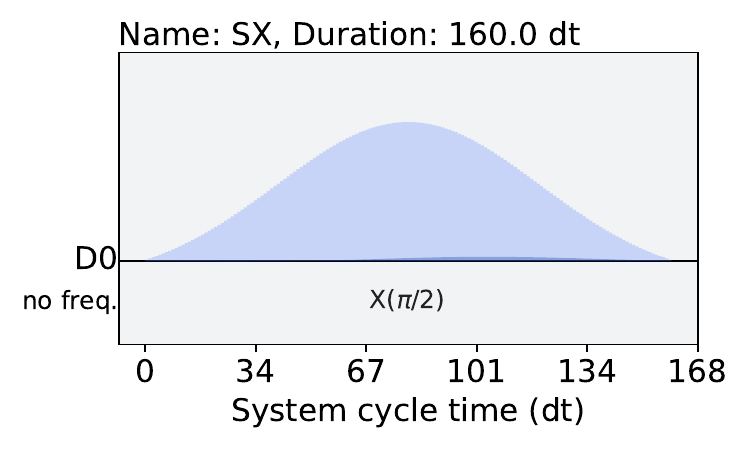}
    \caption{\footnotesize {\tt SX} pulse}
  \end{subfigure}
  \caption{Example {\tt SX}, {\tt X}, and {\tt CX} gate control pulses. Figure from~\cite{xu2023exploration}.}
  \label{fig_basis_gates_pulses_example}
\end{figure}

Microwave pulses are typically used to control superconducting qubits. The right control pulses corresponding to each basis gate must be generated and supplied to the quantum computer in order for it to execute each basis gate. Figure~\ref{fig_basis_gates_pulses_example} displays examples of control pulses for the {\tt SX}, {\tt X}, and {\tt CX} gates. The {\tt I} gate on IBM Quantum has no effect and is effectively just a delay between pulses. In addition, the {\tt RZ} gate is a virtual gate without an actual pulse.

Typically, the envelope, frequency, and phase together characterize a pulse. In the case of the superconducting qubit control, the frequency and phase specify the carrier signal that is to be modulated by the lower-frequency envelope signal. The local oscillator (LO) generates the low phase-noise microwave carrier signal. The envelope specifies the shape of the signal that is created by the arbitrary waveform generator (AWG). The envelop signal is mixed with the carrier signal and that is transmitted to the qubit or couplings to drive operation of the quantum computer. Figure~\ref{fig_hardware_sketch} displays the standard devices for driving the qubits.

Despite the fact that envelopes can have any design, they are often parameterized by a few preset forms, requiring a minimal number of parameters to specify the envelope. These factors often include duration, which indicates how long the pulse is, amplitude, which indicates how strong the pulse is, and other parameters, which determine the pulse's structure.  For instance, the \textit{Derivative Removal by Adiabatic Gate (DRAG) pulse}, which is defined by sigma, which specifies how wide or narrow the Gaussian peak is, and beta, which specifies the correction amplitude, as well as the duration and amplitude, is a standard Gaussian pulse with an additional Gaussian derivative component and lifting applied. Another illustration is the \textit{Gaussian square pulse}, which is a square pulse with a rise-fall in the shape of a Gaussian on each side that has been raised such that its initial sample is zero. It is parameterized by sigma, which determines the width of the Gaussian rise-fall, the width of the embedded square pulse, and the ratio of the duration of each rise-fall to sigma, in addition to the duration and amplitude.

All native gates on IBM Quantum have predetermined pulses, and calibrations are used periodically to adjust their parameters so they can continue to operate with high fidelity over time.

\subsection{Pulse-Level Circuit Description}

To completely define a quantum circuit, all necessary pulses must be specified, together with their timing in relation to the circuits' beginning point and the qubits to which they will be applied. A sequence of pulses (each defined by envelope, frequency, and phase) and the qubits or couplings they operate on effectively defines a so-called {\em pulse-level circuit description}. The superconducting quantum computer control equipment generates and delivers the pulses through RF cables to the cryogenic fridge wherein the qubits and couplings are located.

\subsection{Running Circuits on Quantum Computers}

\begin{figure*}[t]
  \centering
  \includegraphics[width=1\columnwidth]{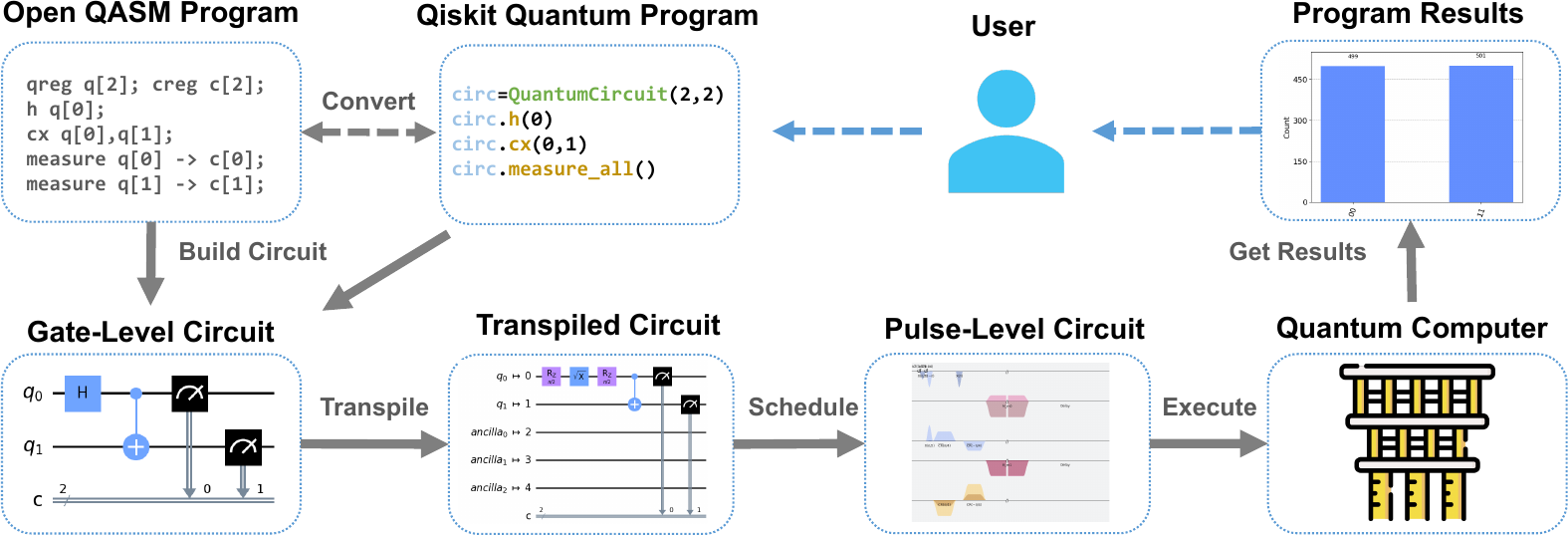}
  \caption{Example of process for running quantum circuits on superconducting quantum computers in Qiskit.}
  \label{fig:qiskit_process}
\end{figure*}

In Figure~\ref{fig:qiskit_process}, we demonstrate a typical Qiskit example on IBM Quantum for running quantum circuits.
The quantum circuits are typically represented as code, as shown by QASM Circuit Specification or Qiskit Circuit Specification in Figure~\ref{fig:qiskit_process}. Similar to classical computing, quantum circuits typically consist of complex instructions. The preparation, compilation, and assembly processes used for classical computing programs are analogous to the activities needed to convert quantum circuits into low-level and hardware-specific instructions before they can actually be executed on quantum computers. To be more precise, while there can be infinite ways to describe a quantum circuit with the same goal, ultimately only the native gates that are supported by the quantum computer need to be~used.

As a result, typically the input circuit specification is translated into a ``Gate-Level Circuit", as shown in Figure~\ref{fig:qiskit_process}. Gate-level circuits can be visualized as shown in the figure, where the gate operations are represented by the symbols on the lines and qubits are represented by the lines that go from left to right. Without more information, it is usually assumed that qubits are in the $\ket 0$ state at the beginning of the quantum circuit. Qubits then evolve through left-to-right sequential processes and are controlled by quantum or classical operations specified in the circuit plot. In order to measure, collect, and store qubit data in classical memory for upcoming analyses, measurements are often carried out at the conclusion of the quantum circuit.

\begin{figure*}[t]
  \centering
  \includegraphics[width=0.8\linewidth]{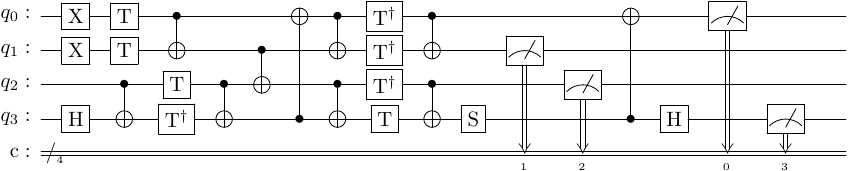}
  \caption{A quantum adder circuit with width=4 (4 qubits) followed by measurement.}
  \label{fig:victim-input}
\end{figure*}

\begin{figure*}[t]
  \centering
  \includegraphics[width=\linewidth]{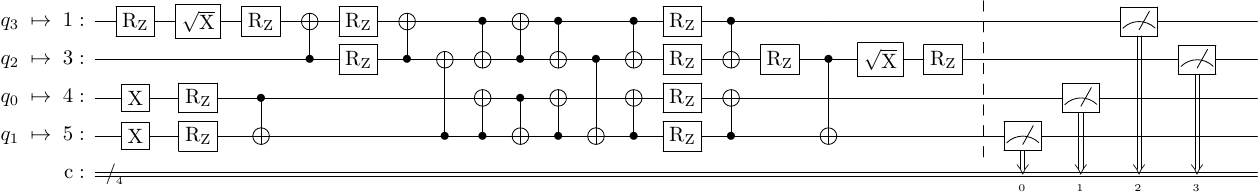}
  \caption{Adder circuit with width=4 transpiled with optimization level 3.}
  \label{fig:victim-circuit}
\end{figure*}

The gate-level circuits are then {\em transpiled}, which is a 
Qiskit term that refers to the operations and transformations that are similar to preprocessing and compilation in classical programs. Transpiling is a multi-step process that involves breaking down non-native quantum gates into groups of native gates, grouping and removing quantum gates to reduce the number of gates, mapping the logic qubits in the original circuits to the physical qubits on the specified quantum computers, routing the circuit under constrained topologies, potentially optimizing circuits to lower error, and more. Following transpilation, circuits are altered in accordance with the knowledge of particular hardware and provide the same logical outcomes as the original circuits.
All of the circuits up to this point are gate-level circuits, which employ a more broad description so that they can be executed in many quantum computers.
Figure~\ref{fig:victim-input} shows one example quantum circuit, and Figure~\ref{fig:victim-circuit} shows one output circuit after transpilation. All the gates are transformed into native gates, and some operations are added to satisfy the topology of the quantum device.

After transpilation, a lower-level procedure occurs, which is known as the {\em schedule} in Qiskit. Microwave pulses, which are the final physical processes needed to regulate and control qubits, are further mapped via scheduling to quantum circuits. Due to scheduling, gate-level circuits are converted into pulse-level circuits. The characteristics that define each microwave pulse—such as amplitude, frequency, and others—were previously covered in Section~\ref{sec::superconducting_quantum_computer_controls}. Scheduling generates microwave pulse sequences based on calibrated data for each basis gate on each qubit or qubit pair and quantum device. The data includes wave envelopes, frequencies, amplitudes, durations, and other characteristics of microwave pulses. All the information that quantum computers require to run the circuit is contained in the final data. This data will be used to alter the qubits of quantum computers once the quantum circuit has begun, and the qubits themselves are controlled by the equipment.

Using the procedures described above, a set of instructions that may be utilized to carry out the required quantum circuits is created from the original quantum circuits. IBM Quantum offers Qiskit as a tool for users to construct circuits, carry out these actions, and submit quantum circuits to the cloud. The cloud will then carry out the users' circuits and execute them before returning the results to users.

%% file: sections/threat_model.tex
\begin{figure*}[t]
  \centering
  \includegraphics[width=0.9\linewidth]{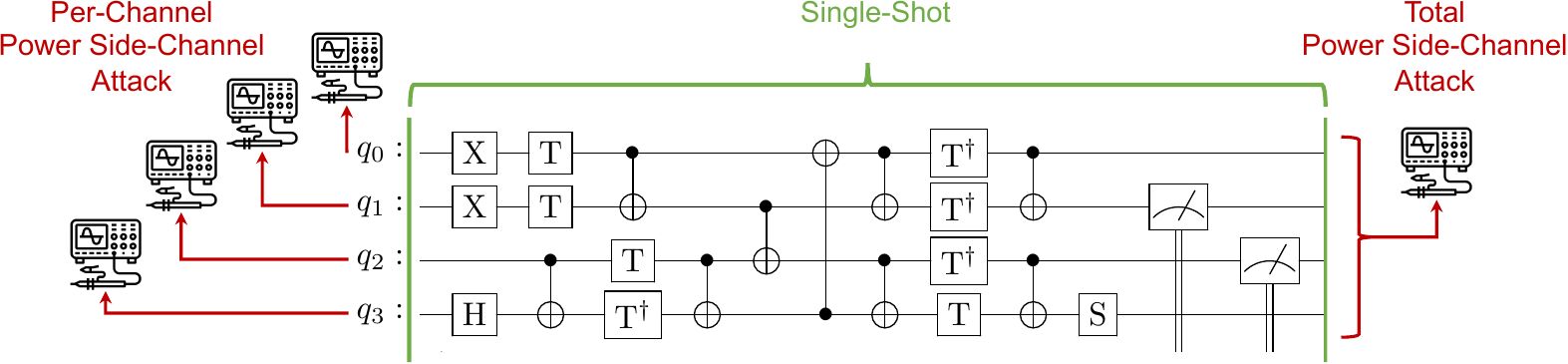}
  \caption{Attack scenarios based on the attacker's measurement capabilities.}
  \label{fig:power-side-channel-attacks}
\end{figure*}

The operation of modern cloud-based quantum computers allows remote users to submit jobs to the cloud provider. These jobs are dispatched to specific quantum computers, also known as backends. While classical computer components, such as the job management server, are considered protected from side-channels, the controller electronics of quantum computers have not been thoroughly analyzed for potential side-channels. The focus of this work is on potential power trace attacks that could extract information about users' quantum circuits from the controllers.

\subsection{Assumptions of Attacker Measurement}

We assume the attacker can sample power traces from shots of a circuit, or they can measure a number of shots and it is easy to divide this into individual shots, since all shots perform the same operations. Recall that each quantum program, i.e. quantum circuit, is executed multiple times, and each execution is called a {\em shot}.

We assume the attacker knows when the victim circuits will be executed so the attacker can capture the side-channel information. Precise knowledge of the execution time is not needed as long as the attacker can capture the trace of one shot. Since the victim often executes thousands of shots, the attacker has multiple chances to capture at least one trace. Each shot is identical without considering the noise.

\paragraph{Single-Shot Per-Channel Power Side-Channel Measurement.}
A stronger attacker is able to collect per-channel power traces (see Figure~\ref{fig:power-side-channel-attacks}). The attacker knows directly which pulses are applied to which qubit as each channel controls different qubits and different two-qubit pairs. Such attackers can attempt Circuit Reconstruction attack from per-channel power traces by collecting a single per-channel power trace for each channel.

\paragraph{Single-Shot Total Power Side-Channel Measurement.}
A weaker attacker could collect a single total power trace over all channels, but not distinguish the power traces of each channel. Such attackers can attempt Circuit Reconstruction attack by collecting a single {\em total} power trace (see Figure~\ref{fig:power-side-channel-attacks}). In particular, there is a trend to have multiple AWGs being part of same physical device. For example, in QICK~\cite{stefanazzi2022qick} framework, FPGAs are used for waveform generation and one FPGA can generate many control pulses. We believe that going forward total power side-channel attacks may be most realistic as attacker may not easily get power traces for individual channel generated by the FPGA, but can easily measure total power consumption of the FPGA, and thus get total power trace of all the channels. Nevertheless, we explore both per-channel and total power side-channel attacks to understand their potential threats.

\subsection{Assumptions of Attacker's Knowledge}

We note that in this work the attacker is assumed to know at all times the information about the target quantum computer (number of qubits it contains, the topology and connections of the qubits) and its basis pulse library. This assumption is reasonable if users have the right to fine-grained control of transpilation and scheduling, because this information is needed in both processes. If this information is not provided, users may easily reverse-engineer it, such as by iteratively increasing the number of qubits to check how many qubits are supported, inserting a two-qubit gate in each qubit pair to check qubit connections, and performing experiments such as frequency sweep and Rabi experiment to acquire the information about the basis pulse library~\cite{xu2023exploration}.

We assume custom gates are not used by users, and all victim circuits are composed only of the basic gates supported by the quantum computer, typically including {\tt ID}, {\tt RZ}, {\tt SX}, {\tt X}, and {\tt CX} for IBM Quantum devices. Among the basic gates, we assume the {\tt RZ} gates are virtual, as is common today. For an attacker who has only access to collect total power traces, we assume he or she knows the in-channel and cross-channel functions that define how the per-channel and total power traces correspond to the pulse information~\cite{xu2023exploration}.

\subsection{Attacker's Objective}

The attacker aims to uncover quantum circuit details from captured power traces. With access to the basis pulse library, which specifies the pulses for all native gates on a specific quantum device, and the measured power traces of the user's circuit, the attacker's goal is to reconstruct the user's circuit. This means retrieving all necessary information about the user's circuit to reproduce it. The attacker seeks to learn the transpiled circuit, which is functionally equivalent to the user's input circuit, even if it may differ in terms of quantum gates used.

\subsection{Impact of Attacks}

Intellectual property, such as quantum algorithm design, is what many users seek to protect. For instance, proprietary quantum machine learning algorithms are being developed by startups who do not own quantum computers; they are worried about the leakage of their proprietary information. Furthermore, different from classical computing, data in quantum computing is encoded as parts of circuits, such as oracles or ansatzes. Besides, input data such as initial states can also be provided eternally to the execution circuits, but it requires quantum memories and quantum networking, which is not available today. As a result, for example, the circuits used sensitive fields, such as medical-related algorithms, may encode private information, and it needs to be protected.

\subsection{The Realism of the Threat Model}
Our work focuses on physical side-channel attacks, such as been widely studied in classical computers. As in classical setting, we assume physical access, which is a standard assumption in any physical side-channel attack. The practicality of these attacks is on the same level as for classical computer power side-channel attacks where attackers can probe the power supply network or power supply of the target (signal generator in our case). Note that attackers can purchase signal generators from science equipment vendors to study their power consumption profile and fine-tune attacks ahead of time.

\subsection{Difference from Classical Setting of Power Side-Channel Attacks}
The major difference of our research is that in classical computers there are no analog control pulses; in classical settings the instructions to the processing unit are digital data read from digital instruction memory, in quantum computers these are analog pulses sent by the signal generators. Our work and threat model assumes any classical and digital information is already protected, and there is a large body of research on the protection of classical computers from power side-channels. Meanwhile, we focus on analog control pulses and signal generators which are not well understood from a security perspective so far.

%% file: sections/formalization.tex
\subsection{Quantum Device}

For a superconducting quantum device, sometimes called a quantum processor, the most important features of the topology are the number of qubits and how they are placed and connected with each other. In addition, each quantum device also has its own native~gates.
In this paper, for a quantum device $D$, we used $n$ to represent its number of qubits, and $m$ to represent its number of qubit connections. The set of basis gates of $D$ is denoted by $BG$. On most of the current quantum devices on IBM Quantum, the basis gates are:
\begin{equation}\label{eq:basis-gates}
    BG = \{ {\tt I, RZ, X, SX, CX} \}.
\end{equation}

\subsection{Channel}
\label{sec_channel_definition}

{\em Channels} refer to which part of the hardware the pulses will be sent to control the qubits. Pulses are applied on one channel for single-qubit gates and several channels for multiple-qubit gates, as described in Section~\ref{sec::superconducting_quantum_computer_controls} on quantum computer controls. The four main categories of channels are drive channels, which send signals to qubits to perform gate operations, control channels, which supplement the drive channel's control over the qubit, measure channels, which send measurement stimulus pulses for readout, and acquire channels, which are used to gather data. Without considering the measurement operations, quantum circuits only trigger drive and control channels. Drive channels typically correspond to qubits, whereas control channels typically correspond to the connections between qubits and are used for two-qubit gates. The architecture of the quantum computer determines how many channels of each type there are.

For a quantum device $D$ we can define a set $C$ to represent the set of channels on $D$. To be more specific, if only considering drive and control channels, $C$ can be represented~as:

\begin{equation}\label{eq:channel_set}
    C = \{drive_0, drive_1, \dots, drive_{n-1}, control_0, control_1, \cdots, control_{m-1}\}
\end{equation}

\noindent where $drive$ refers to the drive channel and $control$ refers to the control channel of $D$, and $n$ and $m$ is the number of qubits and connections of the device $D$.

\subsection{Basis Pulse}\label{sec::basis_pulse}

%The standard procedure for operating a quantum circuit is explained in Section~\ref{sec::background}. 
Every quantum circuit must be translated into a quantum circuit that only includes the target quantum device's basis gates. The group of pulses that follow the scheduling of a basis gate are referred to as its {\em basis pulses}. Because the quantum gate is an abstract notion, pulse parameters for the same type of gate on various channels vary because pulse parameters are highly reliant on qubit physical features. For instance, the pulse parameters of the {\tt X} gate on qubit $0$ are often different from those of the {\tt X} gate on qubits other than $0$. Basis gates and their associated pulse waveform are predetermined, thus they typically do not change over different qubits.
Thus, to define the basis pulse, the gate type as well as the channels need to be specified. We refer this information to labels, and define the set of labels for all the basis gates and possible channels to be:
\begin{equation}\label{eq:label_set}
    L = \{ (gate, C^\prime) | gate \in BG, C^\prime \subset C \}
\end{equation}
where $BG$ is the set of basis gates and $C$ is the set of channels. $L$ represents all basis gates with their channel information, which can uniquely specify a basis pulse.
One basis pulse on the channel $c$ can then be defined as:
\begin{equation}
    p_{l, c}(x) \begin{cases}
        \text{Not always 0} & \text{if } x \in [0, d_l]     \\
        = 0                 & \text{if } x \not\in [0, d_l]
    \end{cases}
\end{equation}
where $c\in l[C^\prime], l \in L$ is the label for the basis pulse and $d_l$ refers to the duration of the pulse, in discrete time steps. The values of $p_{l, c}(x)$ represent the amplitude of the basis pulse with label $l$ at the channel $c$ on time step $x$. All the time steps are in the unit of the system's time resolution, which is denoted as {\tt dt} in Qiskit for IBM Quantum, so the variable $x \in \mathbb{N}$.

Since {\tt I}, {\tt RZ}, {\tt X}, and {\tt SX} are all single-qubit gates, their basis pulses are made up of only one channel. Whereas, {\tt CNOT} gate is a two-qubit gate, so it consists of several channels. For most of the quantum devices on IBM Quantum, the duration of single-qubit gates is chosen to be $160$ {\tt dt}, while the duration of two-qubit gates over different channels is typically different and much longer than the single-qubit gates. For example, $d_{c_i, {\tt X}} = 160$, and $d_{c_j, {\tt CX}} > 1000$ and is often different with different $c_j$.

Because one basis pulse may include pulses on several channels, such as {\tt CX} gate, for each basis gate, its pulses form a set:
\begin{equation}\label{eq:basis_pulse_set}
    p_l(x) = \{p_{l, c}(x) | c \in l[C^\prime]\}.
\end{equation}

\subsection{Basis Pulse Library}
\label{sec::basis_pulse_library}

For all of their quantum devices, IBM Quantum provides the information about basis pulses. We call the collection of basis pulses the {\em basis pulse library}. The so-called {\em custom pulse gates}, which let users produce their own arbitrary pulses, are another feature supported by IBM Quantum, but are left as future work.
Consequently, we assume that there are no custom pulse gates present in the victim circuits. In the end, the basis pulse library can be defined as a set $P_L$ which contains all basis pulses.
\begin{equation}
    \label{eq::basis_pulse_library}
    P_L = \{p_l(x) | l \in L\}.
\end{equation}

\subsection{Pulse-Level Circuit}
In Section~\ref{sec::background}, we mentioned a series of instructions describing how to control the qubits with pulses, making reference to the pulse-level circuit. The circuit's pulse specifications, as well as the start time steps for the instructions, are all contained in the instruction list.
One pulse circuit can be formalized as:
\begin{equation}
    A_{P_L} = \left\{a_{l, t} \cdot p_l (x-t) | l\in L, a_{l, t} \in \{0, 1\}, p_l(x) \in P_L \right\}
    \label{equation::answer}
\end{equation}
where its item $a_{l, t} = 1$ means that there is a basis pulse of label $l \in L$ being applied which starts from the time step $t$, while $a_{l, t} = 0$ means the opposite. As mentioned above, the power traces are discretized in the unit {\tt dt}, and all the time steps are integers, so $t \in \mathbb{N}$. According to {Equation~\eqref{equation::answer}}, $A_{P_L}$ defines all the pulses and where and when they are applied, and thus defines a pulse-level circuit.

\subsection{Power Trace}

The pulses are generated by classical equipment and thus consume energy. The function of the power value with time is what we refer to as the {\em power trace}. The term {\em per-channel power trace} refers to the power trace on a single channel, whereas {\em total power trace} refers to the function of the summation of power over all channels in a time period. Assume that the ability to monitor power consumption on some or all of the channels exists, and that the measured power trace will be made up of and reliant upon a variety of channels. As it reduces multidimensional data to a single dimension, we refer to the function that creates the total power trace from separate power traces as the {\em summation function} or {\em reduction function}  since it reduces multidimensional data to one-dimensional data.

To formalize the power traces, the per-channel power trace and total power trace functions are needed. The per-channel function $Power_c[p_l(x)]$, where $ c \subseteq C$, specifies how the per-channel power traces are computed.
%through the pulse $p_l(x)$. 
The total power trace function $Total[f_c(x)]$, where $ c \subseteq C$, specifies how the total power traces are summed up from all per-channel power traces.
In the experiment, we assume that the per-channel power traces are the square of the norm of the~amplitude:
\begin{equation}
    Power_c[A_{P_L}](x) = \sum_{A_{P_L}} \text{Re}^2[a_{l, t} \cdot p_{l, c}(x - t)] + \text{Im}^2[a_{l, t} \cdot p_{l, c}(x - t)]
\end{equation}
and the total power traces are directly the summation of per-channel power traces:
\begin{equation}
    Total[A_{P_L}](x) = \sum_{c \in C} Power_c[A_{P_L}](x)
    \label{equation:total_power}
\end{equation}

\subsection{Domain-Specific Constraint}
\label{sec:domain_specific_constraints}
In the circuit, we exploit the following constraint (channel constraint): there can only be at most one pulse on each channel at a given time step, which means:
\begin{equation}\label{eq::channel_constraint}
    \begin{aligned}
        \left\{ c_1 = c_2 \land [t_1, t_1 + 1, ..., t_1 + d_{gate_1, c_1}] \cap [t_2, t_2 + 1, ..., t_2 + d_{gate_2, c_2}] = \emptyset \right\} \lor c_1 \neq c_2 \\ \implies 
        \forall t_1\ \text{and}\ t_2 \in [t_1, t_1 + 1, ..., t_1 + d_{gate_1, c_1}],\ a_{c, gate_1, t_1} + a_{c, gate_2, t_2} \in \{0, 1\}
    \end{aligned}
\end{equation}
i.e., if the first pulse with duration $d_{gate_1, c_1}$ has the component on channel $c_1$ starting from time step $t_1$, and the second pulse with duration $d_{gate_2, c_2}$ has the component on channel $c_2$ starting from time step $t_2$, then these two components cannot be mixed with each other.

\subsection{{Attacker's Goal}}

For the per-channel single trace attack, the attacker measures $v_c(x), \forall c\in C$, the per-channel power traces of the victim circuit. For the total power single trace attack, the attacker measures $v(x)$, the total power traces of the victim circuit. The goal of the attacker is to reconstruct the victim circuit, i.e., find a circuit $A_{P_L}$ that is corresponding to the victim circuit. To determine which is better to choose among many circuits, we choose the circuit that minimizes the distance between the power traces with the measured power traces. In addition, the domain-specific constraints discussed in Section~\ref{sec:domain_specific_constraints} need to be~observed.

For the per-channel single trace attack, the goal is:
\begin{equation}
    \label{eq:per_channel_objective}
    A_{P_L} = \argmin_{A^\prime_{P_L}}   \Total_{c\in C}{\left( d\left\{Power_c[A^\prime_{P_L}](x), v_c(x) \right\} \right)}
\end{equation}
where $\Total_{c\in C}(d_c)$ is the function to sum up the distances over all channels to get a total distance. This goal is to find the circuit $A_{P_L}$ from the set of all circuits $\{ A^\prime_{P_L} \}$ that minimizes the distance between the total power traces of this circuit and the measured per-channel power traces.

For the total power single trace attack, the goal is:
\begin{equation}
    \label{eq:total_power_objective}
    A_{P_L} = \argmin_{A^\prime_{P_L}}    d\left\{Total[A^\prime_{P_L}](x), v(x) \right\}
\end{equation}
i.e., finding the circuit $A_{P_L}$ from the set of all circuits $\{ A^\prime_{P_L} \}$ that minimizes the distance between the total power traces of this circuit and the measured total power traces.

%% file: sections/method.tex
In this section, we present two methods that we have developed for stealing quantum program secrets. The first method is based on per-channel single trace information, where the attacker uses per-channel measurements to perform a brute-force attack with the goal of reconstructing the quantum program. The second attack is more challenging, as it restricts the attacker to using only a single total power trace to reconstruct the quantum program. Brute-force methods are not scalable in this case, as the sample pulses at each time step are mixed up, as formulated in Equation~\ref{eq:total_power_objective}. Therefore, we employ Mixed-Integer Linear Programming optimization to find the set of best pulse-level instructions that decompose the quantum program and their corresponding starting time steps.

\begin{figure}[t]
  \small
  \centering
  \caption[A randomly generated circuit on the left is transpiled on a 5-qubit IBM Lima machine.]{A randomly generated circuit on the left is transpiled on a 5-qubit IBM Lima machine (Figure~\ref{fig:lima}). The table on the right shows the starting index of each pulse-level instruction. We aim to recover this table from the measured waveform from each drive channel for the per-channel attacker, or from the total power trace for the total-power side-channel attacker.
  }
  \label{tab:random-circ-complete-reconstruction}
  \raisebox{-3.0ex}{\includegraphics[width=0.35\columnwidth]{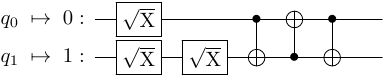}}\quad
  \begin{NiceTabular}{|ll|l|}
    \toprule
    {$\bf dt_{start}$} & \bf{Instruction}        & \small\tt{circ = {\bf QuantumCircuit}(2)} \\
    \midrule
    0                  & {\tt SX}: $\D_1$        & \small\tt{circ.{\bf sx}(1)}               \\
    160                & {\tt SX}: $\D_0$        & \small\tt{circ.{\bf sx}(0)}               \\
    160                & {\tt SX}: $\D_1$        & \small\tt{circ.{\bf sx}(1)}               \\
    320                & {\tt CX}: $\D_0$-$\D_1$ & \small\tt{circ.{\bf cnot}(0, 1)}          \\
    1696               & {\tt CX}: $\D_1$-$\D_0$ & \small\tt{circ.{\bf cnot}(1, 0)}          \\
    3232               & {\tt CX}: $\D_0$-$\D_1$ & \small\tt{circ.{\bf cnot}(0, 1)}          \\
    \bottomrule
  \end{NiceTabular}
\end{figure}

In Figure~\ref{tab:random-circ-complete-reconstruction}, the left part shows a victim circuit (which is a randomly generated for demonstration purposes) that is transpiled on 5-qubit IBM Lima machine (Figure~\ref{fig:lima}). The goal is to recover the circuit from its power trace(s) by finding the set of most suitable pulse-level instructions that make up the circuit with minimum error. The table on the right in Figure~\ref{tab:random-circ-complete-reconstruction} shows the starting time step ($\mathrm{dt_{start}}$) of each pulse-level instruction. If the attacker obtains this information, they can reconstruct the circuit, as the order of the instructions is enough to compile the same circuit again. The third column shows a complete quantum program in Python that is used to generate the circuit on the left. While the per-channel attacker aims to recover this table from the measured waveform from each drive channel, the total power attacker employs only a single mixed, superimposed amplitude samples due the the fact that the attacker does not have access to the individual drive channels.

\subsection{Single-shot Per-Channel Power Side-Channel Attack}
\label{sec:per-channel-attacker}
\input{sections/method_per_channel_attacker.tex}

\subsection{Single-shot Total Power Side-Channel Attack}
\label{sec:total-power-attacker}
\input{sections/method_total_power_attacker.tex}

%% file: sections/method_per_channel_attacker.tex
The process of reconstructing a quantum circuit in the per-channel single trace attack involves measuring the output of each qubit channel and obtaining the waveform of the pulse sequence from each channel. These waveforms are then compared with a set of pulse sequences that include instructions {\tt SX}, {\tt X}, and {\tt CX}, which are obtained from the pulse library of the quantum computer and are unique to each qubit channel. These pulse sequences can be thought of as profiles in classical side-channel analysis, and the most likely instruction profile needs to be identified. To do that, we compare the measured waveform with each candidate profile and calculate the distance between them. However, measurement errors due to noise, or miscalibration of the measuring device can affect the accuracy of the results. Therefore, it is important to have a reliable means of quantifying the similarity between the measured and candidate waveforms.

In Equation~\ref{eq:per_channel_objective}, the objective function is defined as the minimum distance between the measured waveform and the candidate pulse sequences. The distance function $d$ can be naturally defined as the Euclidean distance between two power traces:
\begin{equation}\label{eq:default-distance}
  d_2:\left\{v(x), Total[A_{P_L}](x) \right\} \mapsto\|v(x), Total[A_{P_L}](x)\|_2 = \sqrt{\sum_{i=1}^n\left(v_i(x), Total[A_{P_L}](x)\right)^2}
\end{equation}
where $v_i(x)$ and $Total_{A_{P}}(x)_i$ are the $i$-th elements of $v(x)$ and $Total_{A_{P}}(x)$ respectively. However, we evaluated various distances and metrics on randomly generated circuits including Euclidean distance and found that the \textit{Jensen-Shannon distance} is the most suitable one for quantifying the candidate instructions against the measurements. This distance metric is commonly used to measure the dissimilarity between probability distributions and is well-suited for the task of comparing the probability distributions of pulse sequences.
It has been applied to genome comparison\cite{sims2009alignment, itzkovitz2010overlapping} in protein surface comparison~\cite{ofran2003analysing}, in machine learning~\cite{goodfellow2020generative} and particularly in the analysis of the similarity between two quantum states~\cite{osan2022quantum}.
To compute the Jensen-Shannon distance, we first convert the trace and pulse data of a candidate instruction to two probability vectors $P$ and $Q$ by normalizing the amplitudes to turn them into discrete probability distributions. We then calculate the distance between the two probability vectors by computing the Jensen-Shannon distance (metric) between two probability arrays. This is the square root of the \textit{Jensen-Shannon divergence} ($\sqrt{{\rm JSD}}$)~\cite{endres2003new}.

\textit{Jensen-Shannon divergence} (JSD) is a symmetric, smooth, and bounded measure of dissimilarity between probability distributions that is a well-behaved version of the \textit{Kullback-Leibler divergence} (KLD) $D_{KL}(P \parallel Q)$. It is widely used in information theory and statistics to measure the distinguishablity between probability distributions.
Let $M$ be $\frac{1}{2}(P+Q)$, then the Jensen-Shannon divergence is defined as:
\begin{equation}
  D_{JS}(P \parallel Q)= \frac{1}{2}D_{KL}(P \parallel M)+\frac{1}{2}D_{KL}(Q \parallel M),
\end{equation}

For \textit{discrete probability distributions} $P$ and $Q$ defined on the same sample space, $\mathcal{X}$, the relative entropy from $Q$ to $P$ is defined~\cite{mackay2003information} to be:
\begin{equation}
  D_{KL}(P \parallel Q) = \sum_{x\in\mathcal{X}} P(x) \log\left(\frac{P(x)}{Q(x)}\right).
\end{equation}

Over two probability vectors $x$ and $y$, we compute relative entropy, $D_{KL}(P \parallel Q)$, as an elementwise operation as follows:
\begin{equation}
  D_{KL}(x, y)=\begin{cases}
    x \log(x/y) & x>0, y>0          \\
    0           & x=0, y \geq 0     \\
    \infty      & \text{otherwise }
  \end{cases}
\end{equation}

\begin{table}[t]
  \centering
  \footnotesize
  \caption[Different distance measures for each candidate pulse-level instruction against the measured waveform]{Different distance measures for each candidate pulse-level instruction against the measured waveform whose error rate is 0.1\% while running the circuit in Figure~\ref{tab:random-circ-complete-reconstruction} on Drive Channel(1) on IBM Lima machine (Figure~\ref{fig:lima}). Total duration of the circuit is 4608 dt. RMSE: Root Mean Squared Error, $d_2$: Euclidean distance, $\sqrt{\rm JSD}$: Jensen-Shannon Divergence distance. For starting indices 160, 320, and 4448 dt, all instructions that can be fit in the time window are considered. While at 160 dt and 320 dt, {\tt SX}, {\tt X} and all {\tt CX} instructions are considered, at 448 dt, only {\tt SX} and {\tt X} are considered.
    \textcolor{red}{$\star$} shows the chosen instruction at a $\text{dt}_\text{start}$ whereas {$\circ$} indicates a candidate that is considered but not selected. At $\text{dt}=4448$, {\tt SX} instruction is not selected due to high $\sqrt{\rm JSD}$. \vspace*{1em}}
  \label{tb:complete-reconstruction}
  \begin{NiceTabular}{|clclll|}
    \toprule
    {$\bf dt_{start}$} & {$\bf{qubit_1}$}        & $\bf\checkmark$          & \bf{RMSE}            & ${d_2}$           & $\mathbf{\sqrt{\rm\bf JSD}}$ \\
    \midrule
    $\dots$            & $\dots$                 &                          & $\dots$              & $\dots$           & $\dots$                      \\
    \midrule
    160                & {\tt CX}: $\D_0$-$\D_1$ &                          & 0.00001647           & 0.0128035         & 0.402693706                  \\
    160                & {\tt CX}: $\D_1$-$\D_0$ &                          & 0.00001653           & 0.0147163         & 0.414806197                  \\
    160                & {\tt CX}: $\D_1$-$\D_2$ &                          & 0.00001648           & 0.0157793         & 0.420431628                  \\
    160                & {\tt CX}: $\D_1$-$\D_3$ &                          & 0.00002547           & 0.0190306         & 0.433682782                  \\
    160                & {\tt CX}: $\D_2$-$\D_1$ &                          & 0.00001827           & 0.0146580         & 0.428686046                  \\
    160                & {\tt CX}: $\D_3$-$\D_1$ &                          & 0.00002844           & 0.0172899         & 0.391024599                  \\
    160                & {\tt SX}: $\D_1$        & \textcolor{red}{$\star$} & 0.00000004           & 0.0000906         & 0.000000003                  \\
    160                & {\tt X~}: $\D_1$        &                          & 0.00087999           & 0.0273263         & 0.000014711                  \\
    \midrule
    $\dots$            & $\dots$                 &                          & $\dots$              & $\dots$           & $\dots$                      \\
    \midrule
    320                & {\tt CX}: $\D_0$-$\D_1$ & \textcolor{red}{$\star$} & 0.00000029~$\dagger$ & 0.0014418~$\ddag$ & 0.018015119~$\S$             \\
    320                & {\tt CX}: $\D_1$-$\D_0$ & $\circ$                  & 0.00000016~$\dagger$ & 0.0014530~$\ddag$ & 0.020663924~$\S$             \\
    320                & {\tt CX}: $\D_1$-$\D_2$ &                          & 0.00000404           & 0.0078170         & 0.166908652                  \\
    320                & {\tt CX}: $\D_1$-$\D_3$ &                          & 0.00002301           & 0.0180839         & 0.464903649                  \\
    320                & {\tt CX}: $\D_2$-$\D_1$ &                          & 0.00000105           & 0.0035068         & 0.084267418                  \\
    320                & {\tt CX}: $\D_3$-$\D_1$ &                          & 0.00002377           & 0.0158058         & 0.396308179                  \\
    320                & {\tt SX}: $\D_1$        & $\circ$                  & 0.00000008           & 0.0001265         & 0.000000003                  \\
    320                & {\tt X~}: $\D_1$        &                          & 0.00086605           & 0.0271090         & 0.000014711                  \\
    \midrule
    $\dots$            & $\dots$                 &                          & $\dots$              & $\dots$           & $\dots$                      \\
    \midrule
    4448               & {\tt SX}: $\D_2$        &                          & 0.00025480           & 0.00726239        & 0.450676207                  \\
    4448               & {\tt X~}: $\D_1$        &                          & 0.00133674           & 0.03367971        & 0.450681387                  \\
    \bottomrule
  \end{NiceTabular}
\end{table}

In per-channel single trace attacks, we use the Kullback-Leibler divergence (KLD) and the Jensen-Shannon divergence (JSD) and its distance ($\sqrt{\text{JSD}}$) to distinguish two probability distributions. Therefore, here we provide a brief introduction to divergences, distances and metrics. For more details, we refer the reader to~\cite{deza2009encyclopedia, hayashi2014introduction}.
A \textit{metric} $d$ on a set $\chi$ is a function $d: \chi \times \chi \rightarrow \mathbb{R}_{\geq 0}$ such that for any $x, y, z \in \chi$ the following properties are satisfied:
\begin{align}
  d(x, y) & \geq 0 \label{non-negativity}\tag{non-negativity}                                                 \\
  d(x, y) & =0 \text{ if and only if } x=y \label{idendity-of-indiscernibles}\tag{identity of indiscernibles} \\
  d(x, y) & =d(y, x) \label{symmetry}\tag{symmetry}                                                           \\
  d(x, y) & \leq d(x, z)+d(z, y) \label{triangle-inequality}\tag{triangle inequality}
\end{align}
$\chi$ represents the set of probability distributions and $x$ or $y$ represent an entire probability distribution such as $P=\left\{p_1, p_2, \ldots, p_n\right\}$ where $p_i \geq 0$ and $\forall i, \sum_{i=1}^n p_i=1$. Often, if a distance measure $d$ only satisfies the property~\ref{non-negativity}, is called a \textit{divergence}. If, in addition, $d$ satisfies the properties \ref{idendity-of-indiscernibles}, and \ref{symmetry}, it is called a \textit{distance}. Thus, KLD is a divergence, JSD is a distance, and $\sqrt{{\rm JSD}}$ is a metric~\cite{osan2022quantum}.

\begin{figure}[t]
  \centering
  \includegraphics[trim=18.0 0.0 1 0.0,clip,width=0.75\columnwidth]{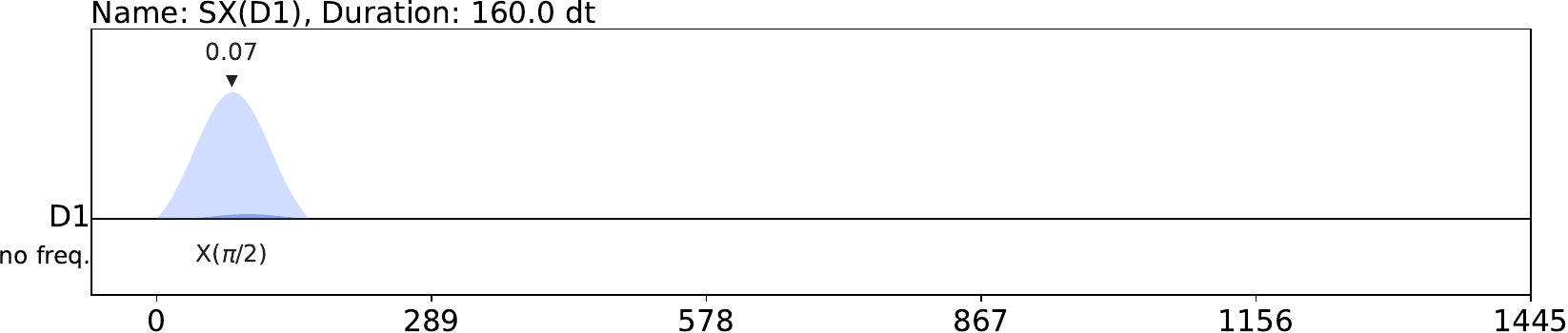}
  \includegraphics[trim=18.0 0.0 1 0.0,clip,width=0.75\columnwidth]{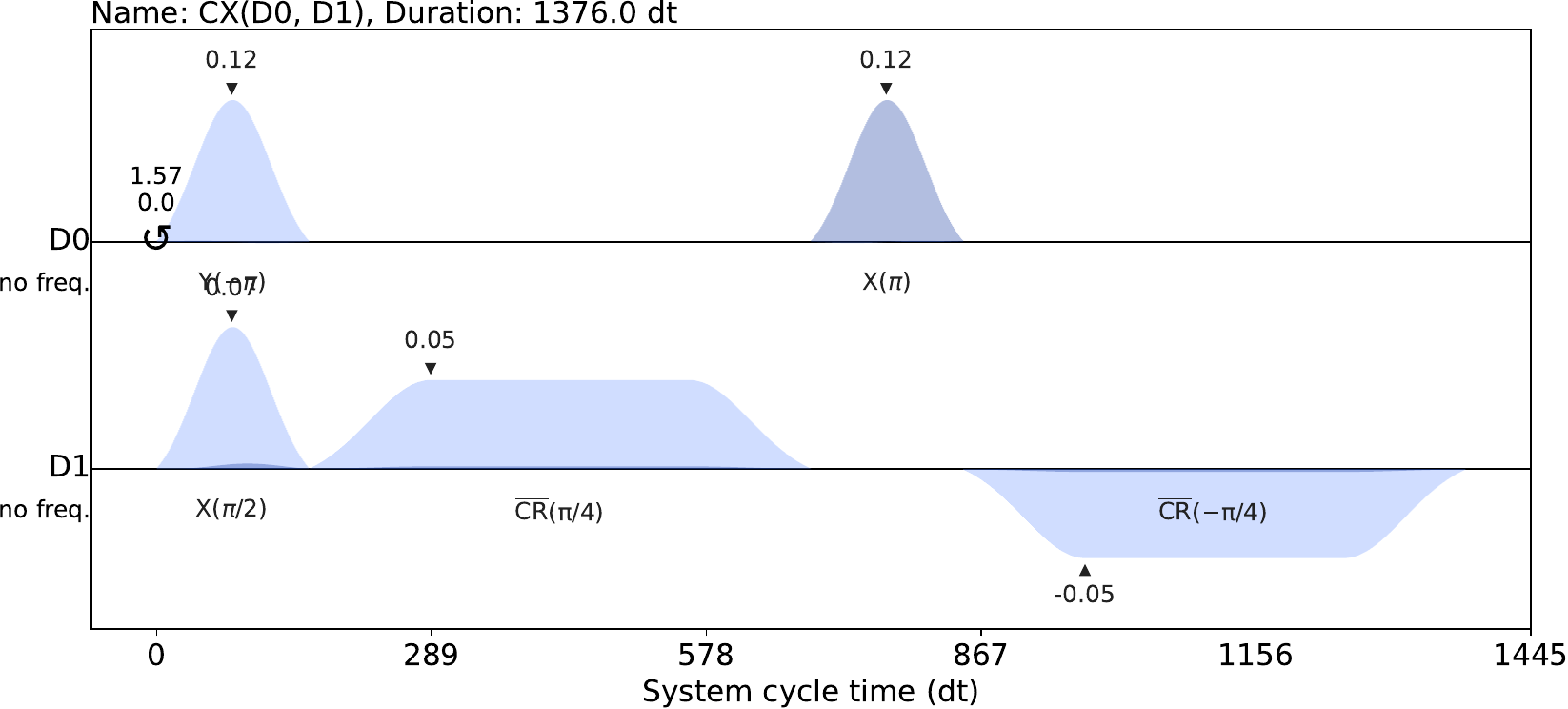}
  \caption[Example pulse schedules]{Example pulse schedules of {\tt SX}: $\D_1$ (\textcolor{red}{$\star$}) and {\tt CX}: $\D_0$-$\D_1$ (\textcolor{black}{$\circ$}), whose durations are 160dt and 1376dt respectively. As seen from the figure, the first waveform on the drive channel $\D_1$ of {\tt CX} is exactly the same pulse as in {\tt SX}.}\label{fig:waveform_examples}
\end{figure}

Figure~\ref{tab:random-circ-complete-reconstruction} is a randomly generated circuit using {\tt SX}, {\tt X}, and {\tt CX} gates. It is transpiled without optimization over {\tt IBM Lima} machine in order to keep the layout of the circuit intact for the sake of the presentation. After transpilation, $q_0$ maps to \textit{Drive Channel 0} and $q_1$ maps to \textit{Drive Channel 1}. The measurement is performed on $q_0$ and $q_1$.

In Table~\ref{tb:complete-reconstruction}, we also show the result of other distance/metrics for comparison and how they perform on the same data. We can see that the Jensen-Shannon distance is the most suitable one for our task. For instance, it is the only one that can distinguish between the three candidate instructions at 320 dt. The other metrics (RMSE and $d_2$) are not able to distinguish between the two candidate instructions shown with $\dagger$ and $\ddag$ in the first two rows at 320dt. However, $\sqrt{\rm JSD}$ is able to distinguish them. Based on low $\sqrt{\rm JSD}$, {\tt SX} gate also looks a good candidate, but we don't select it since as discussed in Figure~\ref{fig:waveform_examples}, its first waveform on the drive channel $\D_1$ of {\tt CX} is {\em always} exactly the same pulse as in {\tt SX}. Therefore, in these cases, {\tt CX} gates always have the priority in selection. At 160 dt, {\tt SX} is chosen since $\sqrt{\rm JSD}$ is considerably smaller than the other two candidates (0.000000003). At 4448 dt, no instruction is selected since {\tt SX} and {\tt X} have very high $\sqrt{\rm JSD}$ values.

We evaluated this metric over the real quantum circuits shown in Table~\ref{tab:benchmarks} and recovered the correct instruction in all cases under different additive Gaussian noise levels where $\mathcal{N}(x, \sigma)$, that is $x$ is the sample amplitude obtained from power trace and $\sigma$ is the standard deviation varies from $0$ (no noise) to $0.1$.

%% file: sections/method_total_power_attacker.tex
\input{sections/background_lira.tex}

\input{sections/background_milp.tex}

\subsubsection{From LIRA to MILP}\label{sec::lira2milp}
{To effectively solve the problem using MILP solvers, we needed to convert the Linear Mixed Integer Real Arithmetic (LIRA) constraints into the MILP form.}
It is worth noting that while SMT solvers can solve the decision variant of the MILP problem, their performance may not be as competitive as dedicated MILP solvers. To illustrate this, we conducted an experiment using the 4-bit adder example from Figure~\ref{fig:victim-circuit}. After an hour of computation, the Z3 SMT solver was still unable to find a satisfiable solution, whereas the same problem was solved in less than 10 seconds using the Gurobi MILP solver. Although there have been efforts to extend SMT solvers to improve their efficiency in solving both decision and optimization variants of the MILP problem~\cite{devriendt2021learn, king2014leveraging}, their performance is still relatively limited. There are also Optimization Modulo Theory (OMT) solvers, such as Z3Opt~\cite{z3opt} and OptiMathSat~\cite{mathsat5}, but these are specialized for finite domains like bitvector theory and are not suitable for our problem {due to the fact that encoding continuous variables as fixed point representation would lead to significant numeric errors in solving real-valued optimization function}. In a recent work on classical power side-channel attacks, these solvers were used to detect vulnerabilities in post-quantum cryptographic primitives~\cite{erata2023towards}. However, their performance was also found to be suboptimal, leading the authors to resort to sampling-based methods for vulnerability identification using bitvector theory in SMT solvers.

The main idea behind MILP solvers is to relax the integer constraints, i.e., treating integer variables as continuous variables initially, and solve the corresponding linear programming (LP) problem.
When the integer constraints are relaxed and integer variables are treated as continuous variables, the solver may encounter rounding errors, precision limitations, or approximation errors in the calculations involving the continuous relaxation of the integer variables. These numerical errors can potentially affect the accuracy and precision of the solution obtained by the solver.
Therefore, over the course of the development of our method, we checked that the optimum configurations returned by the MILP solver are satisfiable by encoding them as decision problems over micro benchmarks (random circuits with a small number of gates) and verifying the satisfiability of the solution using the SMT solver.

In order to encode LIRA constraints into the MILP normal form, {thereby to perform the side-channel attack}, we employed various linearization techniques. {These techniques involve introducing additional binary variables and bounding them with so-called Big-M~\cite{balas1979disjunctive, castro2012generalized} values to convexify non-convex problems that exhibit suitable patterns~\cite{cococcioni2021big}. Overall, we utilized various following linearization techniques from operations research field. In the following sections, we describe these techniques in detail.}

\subsubsection{Linearization of Absolute Valued Objective Function}
\label{sec::linearize-abs}

In order to optimize {Equation~\eqref{eq:total_power_objective}} for a total power single trace attack, we chose to utilize a distance function that can be linearized. {This is one of the most crucial technical insights of our work, paving the way for a complete reconstruction from a single-shot power trace.} For this purpose, we selected the Sum of Absolute Differences (SAD) as our distance function.
\begin{equation}\label{eq:sad}
  d_1:(v(x), Total_{A_{P}}(x)) \mapsto\|v(x), Total_{A_{P}}(x)\|_1 = \sum_{i=1}^n\left|v_i(x)-Total_{A_{P}}(x)_i\right|
\end{equation}
{The $abs$ function is not linear, therefore, it does not allow this metric to directly deal with in the optimization problems, but it can be linearized~\cite{shanno1971linear, lp-solve}}.
An absolute value of a real number can be described as its distance away from zero, or the non-negative magnitude of the number. Thus,
% \begin{equation*}
$|x|= \begin{cases} -x, & \text { if } x<0 \\ \phantom{-}x,  & \text { if } x \geq 0 \end{cases}$.
% \end{equation*}
In our formulation, coefficient signs of the absolute terms are all positive for our minimization problem. Aiming to bound the solution space for the absolute value term with a new variable, $Z$, an equivalent feasible solution can be described by splitting the constraint into two.
If $|X|$ is the absolute value term in our objective function, two additional constraints are added to the linear program: $X \leq Z \land -X \leq Z$.
The $|X|$ term in the objective function is then replaced by $Z$, relaxing the original function into a collection of linear constraints.

\subsubsection{Linearization of Logical Conditions over Binary variables}\label{sec::logical_conditions_over_binary_variables}

In MILP lingo, \textit{binary variables} means decision variables that must take either the value 0 or the value 1, sometimes called 0/1 variables.

{The logical conditions on binary variables and equality relations among binary variables can be converted to binary variables. For instance, we can derive $y= x_1 \lor x_2$ as $y \leq x_1 \land y \leq x_2 \land y \geq x_1+x_2-1$ knowing that  $\lnot x_1$ can be translated into $1 - x_1$ and $x_1 \lor x_2$ into $x_1+x_2 \geq 1$.}

\subsubsection{Encoding Pseudoboolean constraints}\label{sec::pseudoboolean_constraints}
A pseudo-Boolean constraint is an axiom of the form $\sum_i w_i x_i \geq k$, where each $w_i$ and $k$ is a positive integer and each of the $x_i$ is required to have value 0 or 1. In our encoding, we require all weights $w_i$ to be 1, so at least $k$ of $x_1, x_2, \cdots, x_n$ are 1, and at most $k$ of $x_1, x_2, \cdots, x_n$ are 1, and the sum of $x_1, x_2, \cdots, x_n$ is $k$, are all pseudoboolean constraints and can be encoded in MILP form as $x_1 + x_2 + \cdots + x_i \bowtie k$ where $\bowtie \, \in \{\leq,\geq, =\}$ respectively.

\subsubsection{Linearization of Disjunctive Constraints}\label{sec::disjunctive_constraints}
{In order to encode the domain specific constraints given in Equation~\eqref{eq::channel_constraint}, we need to linearize the disjunctive constraints.}

The condition that at least one of the constraints must hold cannot be formulated in a linear programming model, because in a linear program all constraints must hold (conjunction of constraints). In order to solve a disjunctive, the constraints have to be converted into MILP constraints. There are two common methods for disjunction: the Big-M Reformulation and the Convex-Hull Reformulation~\cite{balas1979disjunctive, castro2012generalized}. Here we will only discuss the Big-M Reformulation. Consider the following disjunctive constraint, where $a_i^k$ and $b^k$ are constants, and $x_i$ are variables:
\begin{equation}
  \sum_i  a_i^1 x_i \leq b^1 \lor \sum_i a_i^2 x_i \leq b^2 \lor \sum_i a_i^3 x_i \leq b^3 \lor \cdots \lor \sum_i a_i^k x_i \leq b^k
\end{equation}
For the Big-M reformulation, a sufficiently large number, $M$, is used to nullify one set of constraints. This is accomplished by adding or subtracting the term $M_k*(1-y^k)$ to the upper bound and lower bound constraints, respectively. The bounds are chosen such that they are as tight as possible, while still guaranteeing that the left-hand side of the constraint is always smaller than $b_i+M_i$.
\begin{equation}
  \begin{aligned}
    \sum_i a_i^1 x_i  \leq b^1+M_1\left(1-y^1\right) \land \sum_i a_i^2 x_i  \leq b^2+M_2\left(1-y^2\right) & \land \cdots \\
    y^1+y^2+y^3+\cdots+y^k       \geq 1   \land y^1, y^2, y^3, \cdots, y^k   \in\{0,1\}                     &
  \end{aligned}
\end{equation}
We are able to statically over-approximate $M$ values in our problem since the decision variables $x_i$ are binary variables that can take maximum 1; therefore, for each constraint we basically sum up all the coefficients of the left-hand side of the inequalities, e.g., for the first constraint: $M_1 > \sum_i a_i^j$. To set binary variables $y^j$ to be mutually exclusive, the sum of the variables can be set to 1.

\begin{figure*}[t]
  \centering
  \includegraphics[width=.9\textwidth]{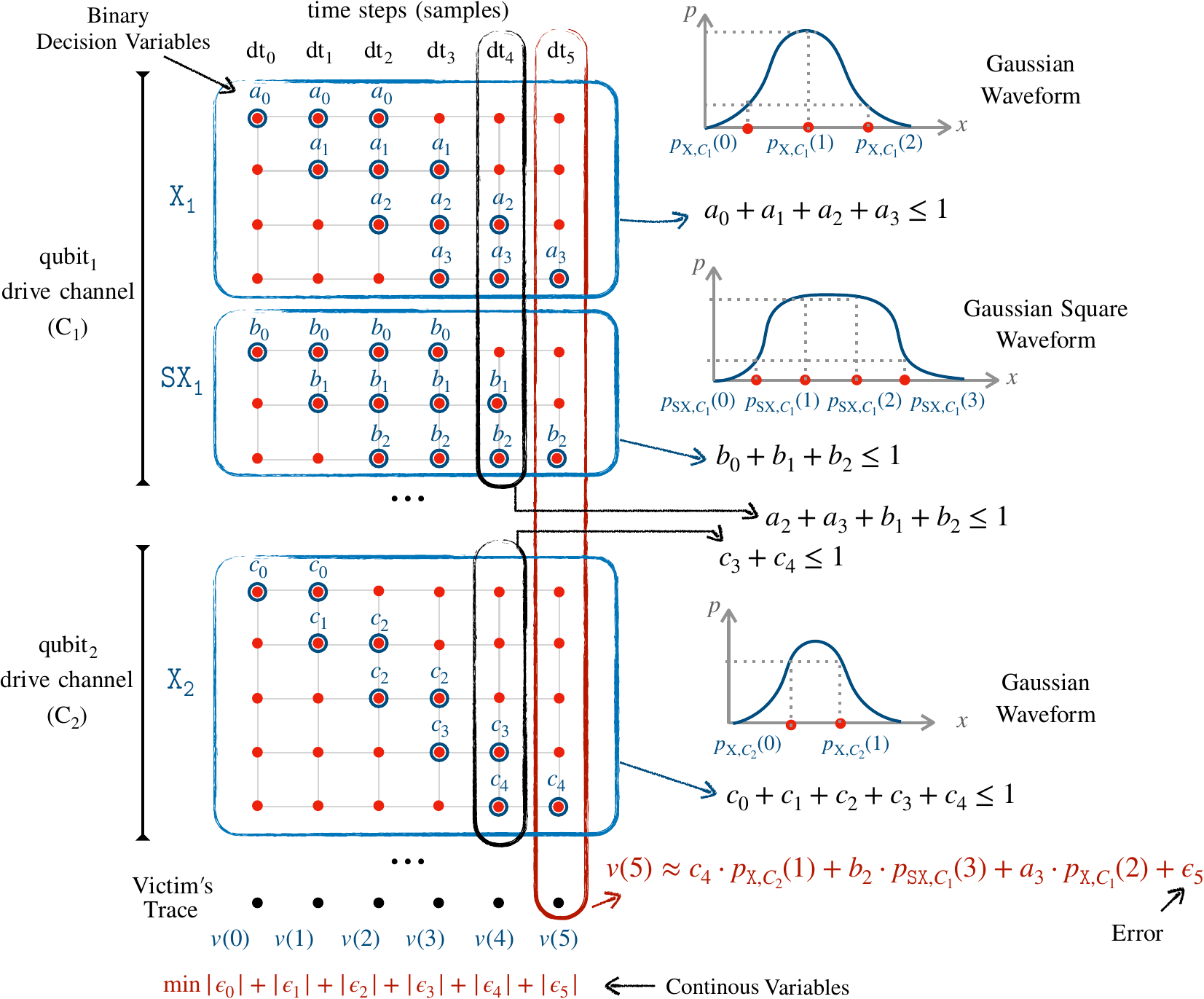}
  \caption{MILP encoding method over an pulse information.}
  \label{fig:milp_encoding_method}
\end{figure*}

\subsubsection{Encoding Decision Variables, Channel Constraints and Objective Function}\label{sec::channel_constraints}
Here we explain an important part of our MILP encoding over an example victim trace sketched in Figure~\ref{fig:milp_encoding_method}. The duration of the trace is 5 $\mathrm{dt}$. The victim runs their circuit on a hypothetical 2-qubit quantum device, $D$, having only two basis gates $BG_D = \{ {\tt X, SX} \}$ ({see Equation~\eqref{eq:basis-gates}}) and two channels $C_D = \{\langle C_1, drive_0\rangle, \langle C_2:drive_1 \rangle\}$ ({see Equation~\eqref{eq:channel_set}}). We simplify the notion of label here since {\tt X} and {\tt SX} gates have only one label and waveform: {\tt X} gate on $C_1$, {\tt SX} gate on $C_1$, and {\tt X} gate on $C_2$ are named as ${\tt X_1}$, ${\tt SX_1}$, and ${\tt X_2}$ respectively. The associated pulse waveforem for Gate ${\tt SX_1}$ is \textit{Gaussian} parameterized by $d_{\tt SX_1}=4$ and therefore we have 4 sampling points: $p_{\mathrm{SX}, C_1}(x)$ where $x \in [0, 3]$ {(see $p_{l,t}$ in Equation~\eqref{eq:basis_pulse_set})}. Gate ${\tt X_1}$ and gate ${\tt X_2}$ can be similarly defined and we sketched their waveforms in the figure.

Binary decision variables encircled by blue rounded rectangles in the figure indicates all cases where a candidate gate is applied on the channel. For example, at time step $\mathrm{dt}_0$, a candidate, gate ${\tt X_1}$, is applied on channel $C_1$, its pulse contributes to  $\mathrm{dt}_1$ and $\mathrm{dt}_2$ since its pulse waveform has a duration of $d_{\tt X_1}=3$. The other possible cases where the gate can start are $\mathrm{dt}_1$, $\mathrm{dt}_2$, and $\mathrm{dt}_3$. However, its pulse cannot start at $\mathrm{dt}_4$ and  $\mathrm{dt}_5$ since the pulse waveform has a duration of $d_{\tt X_1}=3$ and the trace has a duration of 5 $\mathrm{dt}$. We create one binary decision variable to represent each possible case where the gate can start ($a_0$, $a_1$, $a_2$, and $a_3$) {(see $a_{l,t}$ in Equation~\eqref{equation::answer})}. Its pulse can either start at $\mathrm{dt}_0$, $\mathrm{dt}_1$, $\mathrm{dt}_2$, or $\mathrm{dt}_3$ or not at all. Therefore, we can encode the binary decision variables as follows: $a_0 + a_1 + a_2 + a_3 \le 1$. If we follow a similar approach for the other gates, we will have 12 binary decision variables and they are encoded as follows:
$a_0 + a_1 + a_2 + a_3 \le 1 \land b_0 + b_1 + b_2 \le 1 \land c_0 + c_1 + c_2 + c_3 + c_4 \le 1$.

Some decision variables at time step $\mathrm{dt} = 4$ are encircled by black rounded rectangles. It captures the channel constraints on $C_1$ and $C_2$ (see Equation~\eqref{eq::channel_constraint}), i.e., there can only be at most one pulse on each channel at a given time step. This can simply be encoded as follows: $a_2 + a_3 + b_1 + b_2 \le 1 \land c_3 + c_4 \le 1$.

The objective function is encoded in such a way that it minimizes the error between victim's power sample $v(i)$ and the power trace of the pulse waveform $p_{{\tt X}, C_2}(1)$, $p_{{\tt SX}, C_1}(3)$, and $p_{{\tt X}, C_1}(2)$ at time step $i$:
\begin{equation}
  \min |\epsilon_0| + |\epsilon_1| + |\epsilon_2| + |\epsilon_3| + |\epsilon_4| + |\epsilon_5|
\end{equation}
{The linearization of this type of objective functions is explained in Section~\ref{sec::linearize-abs}}.
The following formula states that the power trace of candidate pulses should be equal to the victim's power trace at time step $dt = 5$ with an error of $\epsilon_5$ and some tolerance:
\begin{equation}
  v(5) \approx c_4 \cdot p_{{\tt X}, C_2}(1) + b_2 \cdot  p_{{\tt SX}, C_1}(3) + a_3 \cdot p_{{\tt X}, C_1}(2) + \epsilon_5.
\end{equation}

%% file: sections/background_lira.tex
During the development and testing of the {single-shot total power side-channel attack}, we leveraged SMT (Satisfiability Modulo Theories) solvers and the theory of Linear Mixed Integer Real Arithmetic (LIRA). This allowed us to effectively combine the capabilities of SMT solvers with the expressive power of LIRA to analyze and verify the attack's behavior and to check the correctness of the attack's results obtained from optimization solver. 

To implement the attack, we ultimately encoded the problem using MILP (Mixed Integer Linear Programming) solvers due to their efficiency in handling large-scale optimization problems. However, we encountered a challenge in bridging the expressiveness gap between LIRA and MILP, which is discussed in detail in the following sections. We used various encoding tricks to convert LIRA constraints to MILP constraints, as well as encoding of logical conditions, pseudo-boolean constraints, and disjunctive constraints. These efforts allowed us to successfully apply MILP solvers to our attack and achieve accurate and reliable results.

\subsubsection{Linear Mixed Integer Real Arithmetic (LIRA)}\label{sec::LIRA}
In this section, we provide a brief introduction to SMT solvers and LIRA, and we discuss the disadvantages of using LIRA in the context of our~attack.

LIRA is a theory of linear arithmetic with real and integer variables. Modern SMT solvers such as Z3~\cite{z3}, CVC5~\cite{cvc5}, and MathSAT5~\cite{mathsat5} support LIRA constraints and are equipped with decision procedures for arbitrary boolean combinations (e.g. disjunction and conjunction) of linear constraints. 

LIRA considers the reals and integers as domains for the types of identifiers and constants. For the former domain the problem is polynomial, and for the latter the problem is NP-complete~\cite{kroening2016decision}. As an example, the following is a formula in linear arithmetic: $3x_1 + 2x_2 \leq 5x_3 \, \lor \, 2x_1 - 2x_2 = 0$.

%% file: sections/background_milp.tex
\subsubsection{Mixed Integer Linear Programming (MILP)}\label{sec::milp}

LIRA is a potent formalism that combines linear equalities or inequalities with arbitrary boolean connectors, and allows for mixing integer and real variables. This makes it highly convenient for modeling complex reconstruction problems. However, current SMT solvers lack support for optimization problems, as will be elaborated on in Section~\ref{sec::lira2milp}. In our formalization of the total power single trace attack, we utilize both integer and real variables, and encode the optimization model as Mixed Integer Linear Programming (MILP) problem to address this limitation.

In this section, we briefly introduce MILP. A mathematical optimization problem, or just optimization problem, has the form
\begin{equation} \label{eq:optimization-problem}
  \begin{array}{ll}
    \text{min}   & f_0(x)                                \\
    \text {s.t.} & f_i(x) \leq b_i, \quad i=1, \ldots, m
  \end{array}
\end{equation}
Here the vector $x=\left(x_1, \ldots, x_n\right)$ is the optimization variable of the problem, the function $f_0: \mathbf{R}^n \rightarrow \mathbf{R}$ is the objective function, the functions $f_1, \ldots, f_m: \mathbf{R}^n \rightarrow \mathbf{R}$, $i=1, \ldots, m$, are the (inequality) constraint functions, and the constants $b_1, \ldots, b_m$ are the limits, or bounds, for the constraints. A vector $x^{\star}$ is called optimal, or a solution of the problem, if it has the smallest objective value among all vectors that satisfy the constraints: for any $z$ with $f_1(z) \leq b_1, \ldots, f_m(z) \leq b_m$, we have $f_0(z) \geq f_0\left(x^{\star}\right)$

The optimization problem in Equation~\eqref{eq:optimization-problem} is called a Linear Program (LP) if the objective and constraint functions $f_1, \ldots, f_m$ are linear, i.e., satisfy
\begin{equation} \label{eq:linear}
  f_i(\alpha x+\beta y)=\alpha f_i(x)+\beta f_i(y)
\end{equation}
for all $x, y \in \mathbf{R}^n$ and all $\alpha, \beta \in \mathbf{R}$. If the optimization problem is not linear, it is called a nonlinear program.
Integer Linear Programming (ILP) is an extension of Linear Programming (LP) which allows for variables to take on only integer values, rather than continuous values. This makes ILP useful for solving problems in which the decision variables must be integers, such as scheduling, resource allocation, and network design. ILP is particularly useful for linearizing nonlinear programs, which can be difficult or impossible to solve directly. By introducing additional variables and constraints, nonlinear programs can be transformed into a linear form, making them amenable to solution by LP techniques.

Mixed Integer Linear Programming (MILP) is a more general form of ILP, where some variables are restricted to be integers while others can be continuous. The result is a model that can represent a wider range of real-world optimization problems, making it a powerful technique for solving complex optimization problems. In Mixed Integer Normal Form, every atomic formula is of the form: $a_1 x_1+a_2 x_2+\ldots+a_n x_n \bowtie c$ where $\bowtie \, \in \{=, \leq,\geq\}$ where $x_i$ can be integer or continuous.

ILP and MILP are widely used in many areas such as operations research, computer science, engineering and management science. They are implemented in various commercial optimization software such as CPLEX~\cite{cplex}, Gurobi~\cite{gurobi}, and Xpress~\cite{xpress}.

In addition to these use cases, we apply Mixed Integer Linear Programming (MILP) to the reconstruction of quantum circuits from their power traces. The objective is to reconstruct the original quantum circuit as accurately and correctly as possible from power traces obtained by an attacker. The optimization aims to minimize the discrepancy between the total power consumption of the candidate quantum circuits within the search space and the power trace of the original quantum circuit. This optimization challenge is framed as a MILP problem. The MILP formulation allows for the inclusion of both continuous and integer decision variables (e.g., a binary decision variable to determine whether a {\tt CX} gate exists on drive channels 0 and 1), making it a suitable tool for this problem, which involves both continuous and discrete variables.

Our solution to the MILP problem is obtained by using a commercial solver, Gurobi~\cite{gurobi} with a free academic license.
It is known for its high performance and ability to handle large-scale MILP problems.
However, the resulting encoding can be serialized as an {\tt.lp} file in MILP normal form and can be solved by any MILP solver such as open-source PuLP~\cite{pulp}.

%% file: sections/evaluation.tex
%\subsection{Experimental Setup} 
\label{sec::experimental_setup}

\input{sections/experimental_setup}

\section{Evaluation Results}

The results of our evaluation on benchmark quantum circuits provide compelling evidence of the high accuracy and effectiveness of our techniques in reconstructing quantum circuits.
Table~\ref{tab:recovered} shows that we are able to recover all the X, SX, and CX gates from all the benchmarks tested.
It is important to note that while our techniques demonstrate high accuracy in reconstructing quantum circuits from power traces, the resulting circuits may not be an exact replica of the original circuit, but rather a semantically equal circuit. This is analogous to classical compiler optimizations where reverse engineering from binary code to the original C code may not yield an exact reconstruction.

Furthermore, our evaluation on benchmark quantum circuits serves as a strong validation of our methods for handling mixed discrete and continuous variables, as well as overcoming challenges associated with leakage occurring over different qubit and control channels.
As evident from Table~\ref{tab:bench_complexity}, our evaluation shows that the complexity of MILP encoding increases as the number of qubits in our benchmarks increases. This is reflected in the higher number of real and integer variables, as well as the total number of constraints. This is primarily due to the fact that a larger number of qubits in a quantum circuit results in a higher number of gates and operations, which in turn leads to a larger number of variables and constraints in the MILP encoding.
Additionally, we found that the duration of the quantum circuit, which represents the length of time over which the power traces are captured, has an impact on the MILP encoding complexity. Specifically, the duration of the quantum circuit affects the length of the real-valued objective constraint in the MILP encoding, as well as the total number of integer variables. These observations highlight the dependence of the MILP encoding complexity on the size of the quantum circuit, including the number of qubits and the duration of the circuit.

Table~\ref{tab:bench_complexity} shows that out total power attack is able to recover small depth quantum circuits in less than about 10 seconds for a moderate size benchmark with 5 and 6 qubits, it takes at most 50 seconds. For larger benchmarks such as hhl it takes about 10 minutes.

\input{sections/evaluation_tbl_recovered.tex}

The results of the evaluation on benchmark quantum circuits demonstrate the high accuracy and effectiveness of our techniques in reconstructing quantum circuits, although it should be noted that we used noiseless traces for the evaluation in total power attack. However, the optimization method employed in our approach is known to be robust to noise since we are searching for the best configuration that minimizes the error, which presents an interesting avenue for future work to investigate the impact of noise in power traces on the accuracy of our technique.

\begin{table*}[t]
  \footnotesize
  \centering
  \caption[MILP encoding complexities and pulse-level recovery results.]{MILP encoding complexities of the benchmarks and pulse-level recovery results.\vspace*{1em}}
  \begin{NiceTabular}{@{}|l|c|c|c|c|c|c|c|c|@{}}
    \toprule
    \multirow{1}{*}{\textbf{Quantum}} & \textbf{Total}  & \textbf{Total} & \textbf{Circ.} & \textbf{Total} & \textbf{Real} & \textbf{Integer} & \textbf{Total}    & \textbf{Solver} \\
    {\textbf{Circuit}}                & \textbf{Qubits} & \textbf{Gates} & \textbf{Depth} & \textbf{dt}    & \textbf{Vars} & \textbf{Vars}    & \textbf{Constrs.} & \textbf{Time}   \\ \midrule

    deutsch                           & 2               & 10             & 7              & 28432          & 58            & 6032             & 336               & 0.13s           \\
    dnn                               & 2               & 306            & 17             & 31696          & 159           & 19080            & 756               & 0.51s           \\
    grover                            & 2               & 15             & 12             & 30000          & 106           & 11872            & 536               & 0.30s           \\
    iswap                             & 2               & 14             & 9              & 30096          & 111           & 12432            & 556               & 0.37s           \\
    quantumwalks                      & 2               & 38             & 17             & 31600          & 156           & 39000            & 874               & 0.51s           \\ \midrule
    basis\_change                     & 3               & 85             & 45             & 43440          & 526           & 117824           & 2328              & 4.58s           \\
    fredkin                           & 3               & 31             & 32             & 48720          & 691           & 136818           & 2962              & 6.72s           \\
    linearsolver                      & 3               & 26             & 14             & 32688          & 190           & 23370            & 883               & 0.99s           \\
    qaoa\_n3                          & 3               & 35             & 20             & 14464          & 412           & 64272            & 1804              & 4.40s           \\
    teleportation                     & 3               & 12             & 9              & 29648          & 95            & 10545            & 491               & 0.24s           \\
    toffoli                           & 3               & 24             & 20             & 39312          & 397           & 61535            & 1743              & 3.15s           \\
    wstate                            & 3               & 47             & 34             & 50896          & 759           & 157872           & 3244              & 6.73s           \\ \midrule
    adder                             & 4               & 33             & 28             & 57808          & 975           & 282750           & 4190              & 13.3s           \\
    basis\_trotter                    & 4               & 2353           & 469            & 221248         & 1361          & 1503905          & 2722              & 150.2s          \\
    bell                              & 4               & 53             & 18             & 33680          & 221           & 39780            & 1064              & 2.27s           \\
    cat\_state                        & 4               & 6              & 7              & 32688          & 190           & 32870            & 933               & 1.12s           \\
    hs4                               & 4               & 28             & 12             & 31600          & 156           & 26520            & 794               & 2.0s            \\
    inverseqft                        & 4               & 30             & 22             & 26928          & 10            & 200              & 60                & 0.04s           \\
    qft                               & 4               & 50             & 40             & 57072          & 952           & 273224           & 4095              & 17.97s          \\
    qrng                              & 4               & 12             & 4              & 26768          & 5             & 50               & 30                & 0.01s           \\
    variational                       & 4               & 58             & 30             & 41200          & 456           & 97128            & 2037              & 3.92s           \\
    vqe                               & 4               & 78             & 24             & 14624          & 346           & 161928           & 1852              & 5.47s           \\
    vqe\_uccsd                        & 4               & 238            & 198            & 104128         & 1876          & 1130000          & 3752              & 53.91           \\ \midrule
    error\_cd3                        & 5               & 249            & 161            & 96672          & 4017          & 584577           & 6042              & 98.43           \\
    lpn                               & 5               & 17             & 9              & 30352          & 117           & 23634            & 670               & 0.36s           \\
    pea                               & 5               & 126            & 64             & 73744          & 1473          & 720297           & 6381              & 40.34s          \\
    qec\_en                           & 5               & 52             & 23             & 49072          & 702           & 262548           & 3182              & 12.08s          \\
    qec\_sm                           & 5               & 8              & 19             & 60000          & 600           & 189600           & 2716              & 19.11s          \\ \midrule
    qaoa\_n6                          & 6               & 408            & 109            & 70464          & 2202          & 1212456          & 4404              & 162.27          \\
    simon                             & 6               & 65             & 54             & 62976          & 1186          & 849176           & 5460              & 40.95s          \\
    vqe\_uccsd                        & 6               & 3865           & 2883           & 2009248        & 62800         & 32872200         & 125600            & 254s            \\ \midrule
    hhl                               & 7               & 565            & 380            & 265088         & 8288          & 5216264          & 16576             & 136.1s          \\ \bottomrule
  \end{NiceTabular}
  \label{tab:bench_complexity}
\end{table*}

%% file: sections/experimental_setup.tex
\begin{figure*}[t!]
  \captionsetup[subfigure]{justification=centering}
  \centering
  \begin{subfigure}[t]{0.31\textwidth}
    \centering
    \includegraphics[width=0.6\textwidth]{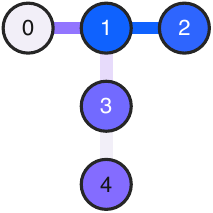}
    \caption{ibmq\_lima (Falcon r4T)\\\small\{CX, ID, RZ, SX, X\}}
    \label{fig:lima}
  \end{subfigure}%
  ~
  \begin{subfigure}[t]{0.31\textwidth}
    \centering
    \includegraphics[width=0.6\textwidth]{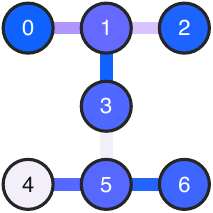}
    \caption{ibm\_lagos (Falcon r5.11H)\\\small\{CX, ID, RZ, SX, X\}}
    \label{fig:lagos}
  \end{subfigure}
  ~
  \begin{subfigure}[t]{0.34\textwidth}
    \centering
    \includegraphics[width=0.9\textwidth]{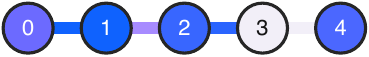}
    \caption{ibmq\_manila (Falcon r5.11L)\\\small\{CX, ID, RZ, SX, X\}}
    \label{fig:manila}
  \end{subfigure}
  \caption[IBM Quantum computers used in the evaluation]{IBM Quantum computers used in the evaluation. Figure shows the device coupling maps. The color of nodes implies frequency (GHz, darker color means lower frequency) of the qubit. The connection color implies the gate time in nanoseconds for 2-qubit gates such as CX (darker color means shorter time).}
  \label{fig:coupling_map}
\end{figure*}

The information about the quantum computer pulses is taken from real quantum computers from IBM Quantum's basic pulse information. For example, we use 7-qubit H-shape superconducting quantum computer, {\tt ibm\_lagos}, (coupling map is shown in Figure~\ref{fig:lagos}) for transpilation and scheduling for all benchmarks. The {\tt ibm\_lagos} is the largest computer we have access to, still, due to the limitation of the number of qubits of even this computer, we chose all algorithms whose numbers of qubits are less or equal to 7. Since pulse parameters for the same type of gate on different machines vary (see Section~\ref{sec::basis_pulse}) due to device topology, for quantum circuit benchmarks up to 5-qubit, we performed additional evaluation on the 5-qubit T-shape {\tt ibm\_lima} and L-shape {\tt ibm\_manilla} machines (their coupling maps are shown in Figure~\ref{fig:lima} and Figure~\ref{fig:manila} respectively).

For the quantum circuits tested, we use well-known benchmarks, listed in Table~\ref{tab:benchmarks}. In particular, we used  QASMBench Benchmark Suite version 1.4\footnote{\url{https://github.com/pnnl/QASMBench/commit/0ae7a5ad97fcc1384230df2b11b8b0ef65ada256}}~\cite{qasmbench} for NISQ evaluation.
QASMBench is a low-level benchmark suite based on the OpenQASM assembly-level intermediate representation (IR)~\cite{openqasm}. It collects commonly used quantum algorithms and routines (e.g., the adder circuit in Figure~\ref{fig:victim-input}) from a variety of distinct domains, including quantum chemistry, simulation, linear algebra, searching, optimization, arithmetic, machine learning, fault tolerance, cryptography, and so on. The benchmark suite covers a wide range of quantum circuits with varying circuit depth and width (i.e., number of qubits).

We removed benchmarks ``ipea'' (iterative phase estimation algorithm) and ``shor'' (Shor's algorithm) for evaluation because they have {\tt Reset} and middle measurement that cannot be scheduled on {\tt ibm\_lagos} due to lack of basis pulses. Unless otherwise specified, we used {\tt seed\_transpiler = 0} to control the randomness and other default parameters for transpilation.
We ran the experiments (power side channel attack using MILP encoding) on an Apple M1 Pro machine with 32 GB of RAM. The Gurobi solver used up to 10 cores.

\begin{table}
  \begin{minipage}{\textwidth}
    \centering
    \footnotesize
    % \footnotesize
    % \scriptsize
    \setlength{\tabcolsep}{0.25em}
    \caption{QASMBench benchmark suite version 1.4~\cite{qasmbench}\vspace*{1em}.}
    \label{tab:benchmarks}
    \begin{NiceTabular}{@{}|llr|c|llr|@{}}
      \toprule
      {\bf Benchmark} & {\bf Algorithm}   & {\bf Reference}      & ~ & {\bf Benchmark} & {\bf Algorithm}  & {\bf Reference}     \\
      \midrule
      deutsch         & Hidden Subgroup   & \cite{openqasm}      &   & hs4             & Hidden Subgroup  & \cite{scaffold}     \\
      iswap           & Logical Operation & \cite{openqasm}      &   & bell            & Logic Operation  & \cite{cirq}         \\
      quantumwalks    & Quantum Walk      & \cite{quantumwalks}  &   & qft             & Hidden Subgroup  & \cite{openqasm}     \\
      grover          & Search            & \cite{agent_anakin}  &   & variational     & Quantum Sim.     & \cite{openfermion}  \\
      ipea            & Hidden Subgroup   & \cite{openqasm}      &   & vqe             & Linear Equation  & \cite{scaffold}     \\
      dnn             & Machine Learn.    & \cite{dnn}           &   & vqe\_uccsd      & Linear Equation  & \cite{scaffold}     \\
      teleportation   & Quantum Comm.     & \cite{teleportation} &   & basis\_trotter  & Quantum Sim.     & \cite{openfermion}  \\
      qaoa\_n3        & Optimization      & \cite{qaoa}          &   & qec\_sm         & Error Correction & \cite{openqasm}     \\
      toffoli         & Logical Operation & \cite{scaffold}      &   & lpn             & Machine Learn.   & \cite{sampaio96}    \\
      linearsolver    & Linear Equation   & \cite{linearsolver}  &   & qec\_en         & Error Correction & \cite{sampaio96}    \\
      fredkin         & Logical Operation & \cite{scaffold}      &   & shor            & Hidden Subgroup  & \cite{ibm_qiskit}   \\
      wstate          & Logical Operation & \cite{openqasm}      &   & pea             & Hidden Subgroup  & \cite{openqasm}     \\
      basis\_change   & Quantum Sim.      & \cite{openfermion}   &   & error\_cd3      & Error Correction & \cite{bench_2017}   \\
      qrng            & Quantum Arith.    & \cite{qrng}          &   & simons          & Hidden Subgroup  & \cite{agent_anakin} \\
      cat\_state      & Logical Operation & \cite{scaffold}      &   & qaoa\_n6        & Optimization     & \cite{cirq}         \\
      inverseqft      & Hidden Subgroup   & \cite{openqasm}      &   & vqe\_uccsd      & Linear Equation  & \cite{scaffold}     \\
      adder           & Quantum Arith.    & \cite{scaffold}      &   & hhl             & Linear Equation  & \cite{ibm_hhl}      \\
      \bottomrule
    \end{NiceTabular}
  \end{minipage}
\end{table}

%% file: sections/evaluation_tbl_recovered.tex
\begin{table}[ht]
  \footnotesize
  \centering
  \caption{Number of Gates, number of RZ gates, number of X, SX, CX gates, and indication which gates were recovered (labeled Rec.) for each quantum program. The circuits are transpiled with {\tt seed\_transpiler = 0}, {\tt optimization\_level = 3}, and other default arguments on {\tt ibm\_lagos}\vspace*{1em}.}
  \begin{NiceTabular}{@{}|l|ccc|c|@{}}
    \toprule

    \multirow{1}{*}{\textbf{Quantum}} & \multirow{1}{*}{\textbf{Total}} & \multirow{1}{*}{{\tt\bf RZ}} & \multirow{1}{*}{{\tt\bf X, SX,}} & \multirow{1}{*}{{\tt\bf X, SX,}} \\

    {\textbf{Circuit}}                & {\textbf{Gates}}                & {\textbf{Gates}}             & {\textbf{CX}}                    & {\textbf{CX}}                    \\

    \midrule
    deutsch                           & 10                              & 6                            & 4                                & $\checkmark$                     \\
    dnn                               & 306                             & 164                          & 142                              & $\checkmark$                     \\
    grover                            & 15                              & 9                            & 6                                & $\checkmark$                     \\
    iswap                             & 14                              & 8                            & 6                                & $\checkmark$                     \\
    quantumwalks                      & 38                              & 20                           & 18                               & $\checkmark$                     \\
    basis\_change                     & 85                              & 43                           & 42                               & $\checkmark$                     \\
    fredkin                           & 31                              & 10                           & 21                               & $\checkmark$                     \\
    linearsolver                      & 26                              & 13                           & 13                               & $\checkmark$                     \\
    qaoa\_n3                          & 35                              & 17                           & 18                               & $\checkmark$                     \\
    teleportation                     & 12                              & 6                            & 6                                & $\checkmark$                     \\
    toffoli                           & 24                              & 11                           & 13                               & $\checkmark$                     \\
    wstate                            & 47                              & 17                           & 30                               & $\checkmark$                     \\
    adder                             & 33                              & 13                           & 20                               & $\checkmark$                     \\
    basis\_trotter                    & 2353                            & 1219                         & 1134                             & $\checkmark$                     \\
    bell                              & 53                              & 29                           & 24                               & $\checkmark$                     \\
    cat\_state                        & 6                               & 2                            & 4                                & $\checkmark$                     \\
    hs4                               & 28                              & 16                           & 12                               & $\checkmark$                     \\
    inverseqft                        & 30                              & 22                           & 8                                & $\checkmark$                     \\
    qft                               & 50                              & 26                           & 24                               & $\checkmark$                     \\
    qrng                              & 12                              & 8                            & 4                                & $\checkmark$                     \\
    variational\phantom{----}         & 58                              & 32                           & 26                               & $\checkmark$                     \\
    vqe                               & 73                              & 32                           & 41                               & $\checkmark$                     \\
    vqe\_uccsd                        & 238                             & 104                          & 134                              & $\checkmark$                     \\
    error\_cd3                        & 249                             & 126                          & 123                              & $\checkmark$                     \\
    lpn                               & 17                              & 10                           & 7                                & $\checkmark$                     \\
    pea                               & 126                             & 61                           & 65                               & $\checkmark$                     \\
    qec\_en                           & 52                              & 24                           & 28                               & $\checkmark$                     \\
    qec\_sm                           & 8                               & 0                            & 8                                & $\checkmark$                     \\
    qaoa\_n6                          & 408                             & 196                          & 212                              & $\checkmark$                     \\
    simon                             & 65                              & 29                           & 36                               & $\checkmark$                     \\
    vqe\_uccsd                        & 2289                            & 769                          & 1520                             & $\checkmark$                     \\
    hhl                               & 1092                            & 461                          & 631                              & $\checkmark$                     \\
    \bottomrule
  \end{NiceTabular}
  \label{tab:recovered}
\end{table}

%% file: sections/future_work.tex
In this section, we discuss future research direction and the main problems that this work leaves open. We also discuss the portability of the general approach to other quantum computer architectures.

\subsection{Application to Future Quantum Computers}

Though this study mainly focuses on superconducting quantum computers, and the evaluations are done on specific IBM machines, the study can be extended to superconducting quantum computers provided by other companies, as well as quantum computers with other qubit technologies. Most of the superconducting quantum computers use the same settings as IBM quantum computers, i.e. the qubits are controlled by microwave pulses generated by quantum computer controllers. Though different superconducting quantum computers typically have different pulses, if the pulse information is available, this work can be applied to them. For example, superconducting quantum computers provided by Rigetti~\cite{rigetti} and Oxford Quantum Circuits~\cite{oqc} through Amazon Braket could be analyzed in same way as IBM's machine. However, for other types of qubit technologies, this study needs to be altered to different types of control signals they use. For example, trapper-ion quantum computers are controlled by both control signals and opto-electronics as shown in~\cite{pino2021demonstration}. While control signals are electromagnetic waves similar to superconducting quantum computer control pulses, there may be other operations that are not existent in superconducting quantum computers, and thus the analysis needs to be changed based on the architecture of such quantum computers.

\subsection{Scalability of the Attack for Future Quantum Computers}

During our experimentation phase, we were constrained by the availability of quantum machines with a maximum of 7 qubits. We expect, however, that with a refined trace cutting strategy, it is feasible to handle much larger quantum circuits encompassing more qubits. In parallel, there is active research on circuit cutting where larger circuits are decomposed into smaller circuits, and our approach naturally can be used against the smaller circuits used in circuit cutting.

\subsection{Challenges in Current Threat Model}

Our current threat model makes certain assumptions, such as the negligible power consumption or timing differences in the AWG or FPGA during the computation of the virtual {\tt RZ} gate. If {\tt RZ} gates or their angles can be discerned from the power traces, it could empower even stronger attackers. Additionally, the unique features of different quantum circuits, like the relative locations and operating qubits of {\tt CX} gates, could be exploited to identify {\tt RZ} gates. Developing heuristics to aid attackers in this endeavor is left an area for future exploration.

\subsection{Potentials in Future Threat Model}

The current landscape does not provide power-related data of control equipment through cloud providers. However, if such data becomes accessible in the future, it could open the door to remote attacks. These attacks could leverage our analysis and pulse recovery techniques without necessitating physical access. Furthermore, other side channels, such as EM or acoustic, might be exploitable from a distance, eliminating the need for direct physical contact.

Our research into power side-channel attacks on quantum computer controllers aims to shed light on potential vulnerabilities that could compromise intellectual property or data security. As quantum computers become more ubiquitous, their susceptibility to physical attacks will likely increase. Drawing from classical security paradigms, we can anticipate a plethora of attack vectors, from EM to optical, especially in quantum computers not based on superconducting qubit technology. Our threat model and explorations serve as a compass, guiding future research in this nascent yet critical domain.

\subsection{Potential Defenses}
High-level ideas of existing side-channel protection techniques can be applied. For example, ideas of {\it randomization} could be used to randomly consume power and confuse the attackers. Or the signal generator could operate in {\it constant power} mode to consume the same power regardless of pulses begin generated. The specific implementation of these defense ideas would be new, as, e.g., means to randomly consume power in a signal generator have not been studied or implemented before from a security perspective. Novely of our work is to point out the threats, so that defenses can be developed.

%% file: sections/related_work.tex
Algebraic cryptanalysis in classical computers involves solving a system of (non-linear) equations over a finite field to recover the secret key of a cryptographic primitive, using inputs and outputs along with known plaintext/ciphertext pairs. One approach is to translate the system of equations into an equivalent satisfiability (SAT) problem instance, which can be solved using a SAT solver such as CryptoMiniSat~\cite{soos2016cryptominisat}. However, the resulting algebraic system and its equivalent SAT problem may not contain enough information for efficient solving of most cryptographic primitives. Additional information, such as side-channel information related to the key, plaintext, or ciphertext, can be used to aid in solving the system, such as physical leakage of intermediate states during encryption or key scheduling. Algebraic side-channel attacks are a type of side-channel analysis which can recover the secret information with a small number of samples~\cite{renauld2010algebraic, mohamed2012improved}. SMT and optimizing pseudo-boolean solvers and other constraint solving techniques are also used in algebraic attacks in classical computers~\cite{kumar2022side, oren2012tolerant, cryptoeprint:2014/683}.
However, in our work, we demonstrated that MILP solvers are considerably faster than SMT solvers for solving systems of linear equations with errors.

In a recent study, Baksi et al.~\cite{baksi2021new} presented an automated analysis of side-channel leakage from software and hardware for stream ciphers using SMT solvers. However, the authors encountered a challenge in using MILP solvers, as their attack required arithmetic operations in addition to Boolean operations. Similarly, in our research against superconducting quantum computers, we also faced a similar challenge and investigated various encoding techniques to overcome this limitation.

Our algebraic side-channel attack against superconducting quantum computers distinguishes itself from previous attacks in several key ways.
First, our attack targets the quantum programs themselves, rather than the classical cryptographic primitives. This introduces unique challenges as the leakage occurs over different channels, akin to parallel computation in classical computers. The signals involved are of mixed {(total)} amplitudes {across all channels; any two amplitudes at any time step might be distributed across any two channels based on the coupling map of the quantum hardware when {\tt CNOT} gates are used in the circuit}. Consequently, formalizing the problem and encoding it into a solvable form requires careful consideration of the qubit and control channels involved, making it a non-trivial task.
Second, our attack is specifically designed to handle real-valued {mixed} amplitudes, further adding to the complexity of the problem.

For superconducting quantum computers, \cite{ash2020analysis} shows that the crosstalk errors could be used in fault injection attacks with malicious circuits, and \cite{deshpande2023design} detects such circuits in quantum programs by expressing both input circuits and malicious circuits as graphs and formulating the their detection as a sub-graph isomorphism finding problem.

Xu et al.~\cite{xu2023exploration, xu2023classification} have proposed {a set of physical attacks} on quantum computers. However, unlike our work, they have not provided a formalization of {total and per-channel power side-channel attacks} and which left the question of reverse engineering of quantum circuits from total power traces is open. Our work focuses on a harder problem that aims to recover quantum gates from a single total power-side channel trace where the problem is NP-hard, however, they attempted to solve a simpler problem that aims to recover the quantum gates from multiple traces considering a powerful attacker, and that can be simply solved with a polynomial time method.

Bell and Tr{\"u}gler~\cite{bell2022reconstructing} have investigated reconstructing quantum circuits on cloud-based superconducting quantum computers. However, the authors did not perform a power side-channel attack. The attack runs a probing circuit before and after a victim circuit and analyzes changes in error rates to make a guess about the victim circuit. This method is applicable in more challenging remote settings, but its capability is limited to distinguishing between only two pre-defined circuits, which must already be known to the attacker.

%% file: sections/conclusion.tex
As the interest in quantum computing grows rapidly, securing quantum programs becomes increasingly important, necessitating thorough analysis of security threats. This paper has presented a novel threat to quantum programs in the form of power side-channel attacks, showcasing the formalization and demonstration of using power traces to {reconstruct quantum circuits}. Through Jensen-Shannon Divergence distance metric and algebraic reconstruction from power traces, two new types of single-trace attacks, per-channel and total {power side-channel} attacks, have been realized. The evaluation on benchmark quantum programs has shown the high accuracy of our techniques in reconstructing quantum circuits. 
Our algebraic side-channel attack distinguishes itself from previous attacks on classical computers in its focus on quantum programs, handling of mixed discrete and continuous variables, and challenges associated with information leakage over different qubit and control channels through {a total power trace}. 
This work underscores the need for further advancements in mitigating such side-channel vulnerabilities in quantum systems to ensure the security of quantum programs in quantum computing environments.
{Future research and development efforts should aim to investigate defense techniques in order to enhance the security of quantum programs in quantum computing environments. We also envision further research on the impact of noise on power traces and exploration of strategies to recover {\tt Rz}~gates.}

%% file: main.bbl
\newcommand{\etalchar}[1]{$^{#1}$}
\begin{thebibliography}{GPAM{\etalchar{+}}20}

\bibitem[ABB{\etalchar{+}}21]{synthesis2021date}
Arnold Abromeit, Florian Bache, Leon~A Becker, Marc Gourjon, Tim G{\"u}neysu, Sabrina Jorn, Amir Moradi, Maximilian Orlt, and Falk Schellenberg.
\newblock Automated masking of software implementations on industrial microcontrollers.
\newblock In {\em 2021 Design, Automation \& Test in Europe Conference \& Exhibition (DATE)}, pages 1006--1011. IEEE, 2021.

\bibitem[ABP19]{synthesis2019tcad}
Giovanni Agosta, Alessandro Barenghi, and Gerardo Pelosi.
\newblock Compiler-based techniques to secure cryptographic embedded software against side-channel attacks.
\newblock {\em IEEE Transactions on Computer-Aided Design of Integrated Circuits and Systems}, 39(8):1550--1554, 2019.

\bibitem[Age19]{agent_anakin}
AgentANAKIN.
\newblock {Grover's Algorithm}.
\newblock \url{https://github.com/AgentANAKIN/Grover-s-Algorithm}, 2019.

\bibitem[{Ama}23]{braket}
{Amazon Web Services}.
\newblock {Amazon Braket}, 2023.

\bibitem[ASAG20]{ash2020analysis}
Abdullah Ash-Saki, Mahabubul Alam, and Swaroop Ghosh.
\newblock Analysis of crosstalk in nisq devices and security implications in multi-programming regime.
\newblock In {\em Proceedings of the ACM/IEEE International Symposium on Low Power Electronics and Design}, pages 25--30, 2020.

\bibitem[Bal79]{balas1979disjunctive}
Egon Balas.
\newblock Disjunctive programming.
\newblock {\em Annals of discrete mathematics}, 5:3--51, 1979.

\bibitem[BBB{\etalchar{+}}22]{cvc5}
Haniel Barbosa, Clark~W. Barrett, Martin Brain, Gereon Kremer, Hanna Lachnitt, Makai Mann, Abdalrhman Mohamed, Mudathir Mohamed, Aina Niemetz, Andres N{\"{o}}tzli, Alex Ozdemir, Mathias Preiner, Andrew Reynolds, Ying Sheng, Cesare Tinelli, and Yoni Zohar.
\newblock cvc5: {A} versatile and industrial-strength {SMT} solver.
\newblock In Dana Fisman and Grigore Rosu, editors, {\em Tools and Algorithms for the Construction and Analysis of Systems - 28th International Conference, {TACAS} 2022}, volume 13243 of {\em Lecture Notes in Computer Science}, pages 415--442. Springer, 2022.

\bibitem[BCHC18]{synthesis2018taco}
Nicolas Belleville, Damien Courouss{\'e}, Karine Heydemann, and Henri-Pierre Charles.
\newblock Automated software protection for the masses against side-channel attacks.
\newblock {\em ACM Transactions on Architecture and Code Optimization (TACO)}, 15(4):1--27, 2018.

\bibitem[BDM{\etalchar{+}}20]{synthesis2020tornado}
Sonia Bela{\"\i}d, Pierre-{\'E}variste Dagand, Darius Mercadier, Matthieu Rivain, and Rapha{\"e}l Wintersdorff.
\newblock Tornado: Automatic generation of probing-secure masked bitsliced implementations.
\newblock In {\em Annual International Conference on the Theory and Applications of Cryptographic Techniques}, pages 311--341. Springer, 2020.

\bibitem[BGKS05]{linearsolver}
Cyril Branciard, Nicolas Gisin, Barbara Kraus, and Valerio Scarani.
\newblock Security of two quantum cryptography protocols using the same four qubit states.
\newblock {\em Physical Review A}, 72(3):032301, 2005.

\bibitem[BKS21]{baksi2021new}
Anubhab Baksi, Satyam Kumar, and Santanu Sarkar.
\newblock A new approach for side channel analysis on stream ciphers and related constructions.
\newblock {\em IEEE Transactions on Computers}, 71(10):2527--2537, 2021.

\bibitem[BPF15]{z3opt}
Nikolaj Bj{\o}rner, Anh-Dung Phan, and Lars Fleckenstein.
\newblock $\nu$z-an optimizing smt solver.
\newblock In {\em International Conference on Tools and Algorithms for the Construction and Analysis of Systems}, pages 194--199. Springer, 2015.

\bibitem[BRN{\etalchar{+}}13]{synthesis2015tc}
Ali~Galip Bayrak, Francesco Regazzoni, David Novo, Philip Brisk, Fran{\c{c}}ois-Xavier Standaert, and Paolo Ienne.
\newblock Automatic application of power analysis countermeasures.
\newblock {\em IEEE Transactions on Computers}, 64(2):329--341, 2013.

\bibitem[BT22]{bell2022reconstructing}
Brennan Bell and Andreas Tr{\"u}gler.
\newblock Reconstructing quantum circuits through side-channel information on cloud-based superconducting quantum computers.
\newblock In {\em 2022 IEEE International Conference on Quantum Computing and Engineering (QCE)}, pages 259--264. IEEE, 2022.

\bibitem[BYT17]{synthesis2017post}
Arthur Blot, Masaki Yamamoto, and Tachio Terauchi.
\newblock Compositional synthesis of leakage resilient programs.
\newblock In {\em International Conference on Principles of Security and Trust}, pages 277--297. Springer, 2017.

\bibitem[CAB{\etalchar{+}}21]{cerezo2021variational}
Marco Cerezo, Andrew Arrasmith, Ryan Babbush, Simon~C Benjamin, Suguru Endo, Keisuke Fujii, Jarrod~R McClean, Kosuke Mitarai, Xiao Yuan, Lukasz Cincio, et~al.
\newblock Variational quantum algorithms.
\newblock {\em Nature Reviews Physics}, 3(9):625--644, 2021.

\bibitem[CBSG17]{openqasm}
Andrew~W Cross, Lev~S Bishop, John~A Smolin, and Jay~M Gambetta.
\newblock Open quantum assembly language.
\newblock {\em arXiv preprint arXiv:1707.03429}, 2017.

\bibitem[CF21]{cococcioni2021big}
Marco Cococcioni and Lorenzo Fiaschi.
\newblock The big-m method with the numerical infinite m.
\newblock {\em Optimization Letters}, 15(7):2455--2468, 2021.

\bibitem[CG12]{castro2012generalized}
Pedro~M Castro and Ignacio~E Grossmann.
\newblock Generalized disjunctive programming as a systematic modeling framework to derive scheduling formulations.
\newblock {\em Industrial \& Engineering Chemistry Research}, 51(16):5781--5792, 2012.

\bibitem[CGSS13]{mathsat5}
Alessandro Cimatti, Alberto Griggio, Bastiaan Schaafsma, and Roberto Sebastiani.
\newblock {The MathSAT5 SMT Solver}.
\newblock In {\em International Conference on Tools and Algorithms for the Construction and Analysis of Systems}, pages 93--107. Springer, 2013.

\bibitem[CK19]{choi2019tutorial}
Jaeho Choi and Joongheon Kim.
\newblock A tutorial on quantum approximate optimization algorithm (qaoa): Fundamentals and applications.
\newblock In {\em 2019 International Conference on Information and Communication Technology Convergence (ICTC)}, pages 138--142. IEEE, 2019.

\bibitem[Cor23]{cplex}
IBM Corporation.
\newblock Ibm ilog cplex optimization studio, 2023.

\bibitem[DBE95]{doi:10.1098/rspa.1995.0065}
David~Elieser Deutsch, Adriano Barenco, and Artur Ekert.
\newblock Universality in quantum computation.
\newblock {\em Proceedings of the Royal Society of London. Series A: Mathematical and Physical Sciences}, 449(1937):669--677, 1995.

\bibitem[DDDD09]{deza2009encyclopedia}
Elena Deza, Michel~Marie Deza, Michel~Marie Deza, and Elena Deza.
\newblock {\em Encyclopedia of distances}.
\newblock Springer, 2009.

\bibitem[Dev22]{cirq}
Cirq Developers.
\newblock Cirq, December 2022.
\newblock {See full list of authors on Github: https://github .com/quantumlib/Cirq/graphs/contributors}.

\bibitem[DGN21]{devriendt2021learn}
Jo~Devriendt, Ambros Gleixner, and Jakob Nordstr{\"o}m.
\newblock Learn to relax: Integrating 0-1 integer linear programming with pseudo-boolean conflict-driven search.
\newblock {\em Constraints}, 26(1-4):26--55, 2021.

\bibitem[DMB08]{z3}
Leonardo De~Moura and Nikolaj Bj{\o}rner.
\newblock Z3: An efficient smt solver.
\newblock In {\em International conference on Tools and Algorithms for the Construction and Analysis of Systems}, pages 337--340. Springer, 2008.

\bibitem[DXT{\etalchar{+}}23]{deshpande2023design}
Sanjay Deshpande, Chuanqi Xu, Theodoros Trochatos, Hanrui Wang, Ferhat Erata, Song Han, Yongshan Ding, and Jakub Szefer.
\newblock Design of quantum computer antivirus.
\newblock In {\em 2023 IEEE International Symposium on Hardware Oriented Security and Trust (HOST)}, pages 260--270. IEEE, 2023.

\bibitem[EPMS23]{erata2023towards}
Ferhat Erata, Ruzica Piskac, Victor Mateu, and Jakub Szefer.
\newblock Towards automated detection of single-trace side-channel vulnerabilities in constant-time cryptographic code, 2023.

\bibitem[ES03]{endres2003new}
Dominik~Maria Endres and Johannes~E Schindelin.
\newblock A new metric for probability distributions.
\newblock {\em IEEE Transactions on Information theory}, 49(7):1858--1860, 2003.

\bibitem[EW14]{synthesis2014cav}
Hassan Eldib and Chao Wang.
\newblock Synthesis of masking countermeasures against side channel attacks.
\newblock In {\em International Conference on Computer Aided Verification}, pages 114--130. Springer, 2014.

\bibitem[Fed16]{teleportation}
Sergueï Fedortchenko.
\newblock A quantum teleportation experiment for undergraduate students, 2016.

\bibitem[FIC23]{xpress}
FICO.
\newblock Xpress optimization suite, 2023.

\bibitem[GO23]{gurobi}
LLC Gurobi~Optimization.
\newblock Gurobi optimizer, 2023.

\bibitem[GPAM{\etalchar{+}}20]{goodfellow2020generative}
Ian Goodfellow, Jean Pouget-Abadie, Mehdi Mirza, Bing Xu, David Warde-Farley, Sherjil Ozair, Aaron Courville, and Yoshua Bengio.
\newblock Generative adversarial networks.
\newblock {\em Communications of the ACM}, 63(11):139--144, 2020.

\bibitem[HIK{\etalchar{+}}14]{hayashi2014introduction}
Masahito Hayashi, Satoshi Ishizaka, Akinori Kawachi, Gen Kimura, and Tomohiro Ogawa.
\newblock {\em Introduction to quantum information science}.
\newblock Springer, 2014.

\bibitem[{IBM}23]{ibm_quantum}
{IBM Quantum}, 2023.
\newblock \url{https://quantum-computing.ibm.com/}.

\bibitem[{IBM}nda]{ibm_qiskit}
{IBM}.
\newblock {Qiskit}: Elements for building a quantum future.
\newblock \url{https://github.com/Qiskit/qiskit}, [n.d.].

\bibitem[{IBM}ndb]{ibm_hhl}
{IBM}.
\newblock {Qiskit}: Solving linear systems of equations using hhl.
\newblock \url{https://qiskit.org/textbook/ch-applications/hhl_tutorial.html}, [n.d.].

\bibitem[IHS10]{itzkovitz2010overlapping}
Shalev Itzkovitz, Eran Hodis, and Eran Segal.
\newblock Overlapping codes within protein-coding sequences.
\newblock {\em Genome research}, 20(11):1582--1589, 2010.

\bibitem[Iosnd]{qaoa}
Joseph~T. Iosue.
\newblock {QAOAPython}: The quantum approximate optimization algorithm implemented on cirq, projectq, and qiskit.
\newblock \url{https://github.com/jtiosue/QAOAPython}, [n.d.].

\bibitem[JPK{\etalchar{+}}14]{scaffold}
Ali JavadiAbhari, Shruti Patil, Daniel Kudrow, Jeff Heckey, Alexey Lvov, Frederic~T Chong, and Margaret Martonosi.
\newblock Scaffcc: A framework for compilation and analysis of quantum computing programs.
\newblock In {\em Proceedings of the 11th ACM Conference on Computing Frontiers}, pages 1--10, 2014.

\bibitem[KBT14]{king2014leveraging}
Tim King, Clark Barrett, and Cesare Tinelli.
\newblock Leveraging linear and mixed integer programming for smt.
\newblock In {\em 2014 Formal Methods in Computer-Aided Design (FMCAD)}, pages 139--146. IEEE, 2014.

\bibitem[KDB{\etalchar{+}}22]{kumar2022side}
Satyam Kumar, Vishnu~Asutosh Dasu, Anubhab Baksi, Santanu Sarkar, Dirmanto Jap, Jakub Breier, and Shivam Bhasin.
\newblock Side channel attack on stream ciphers: A three-step approach to state/key recovery.
\newblock {\em IACR Transactions on Cryptographic Hardware and Embedded Systems}, pages 166--191, 2022.

\bibitem[KHF{\etalchar{+}}20]{kocher2020spectre}
Paul Kocher, Jann Horn, Anders Fogh, Daniel Genkin, Daniel Gruss, Werner Haas, Mike Hamburg, Moritz Lipp, Stefan Mangard, Thomas Prescher, et~al.
\newblock Spectre attacks: Exploiting speculative execution.
\newblock {\em Communications of the ACM}, 63(7):93--101, 2020.

\bibitem[KJJ99]{DBLP:conf/crypto/KocherJJ99}
Paul~C. Kocher, Joshua Jaffe, and Benjamin Jun.
\newblock Differential power analysis.
\newblock In Michael~J. Wiener, editor, {\em Advances in Cryptology - {CRYPTO} '99, 19th Annual International Cryptology Conference, Santa Barbara, California, USA, August 15-19, 1999, Proceedings}, volume 1666 of {\em Lecture Notes in Computer Science}, pages 388--397. Springer, 1999.

\bibitem[Koc96]{DBLP:conf/crypto/Kocher96}
Paul~C. Kocher.
\newblock Timing attacks on implementations of diffie-hellman, rsa, dss, and other systems.
\newblock In Neal Koblitz, editor, {\em Advances in Cryptology - {CRYPTO} '96, 16th Annual International Cryptology Conference, Santa Barbara, California, USA, August 18-22, 1996, Proceedings}, volume 1109 of {\em Lecture Notes in Computer Science}, pages 104--113. Springer, 1996.

\bibitem[KS16]{kroening2016decision}
Daniel Kroening and Ofer Strichman.
\newblock {\em Decision procedures}.
\newblock Springer, 2016.

\bibitem[{LP }23]{lp-solve}
{LP Solve}.
\newblock Absolute values.
\newblock \url{http://lpsolve.sourceforge.net/}, 2023.
\newblock Accessed 22 Fe b. 2023.

\bibitem[LSG{\etalchar{+}}20]{lipp2020meltdown}
Moritz Lipp, Michael Schwarz, Daniel Gruss, Thomas Prescher, Werner Haas, Jann Horn, Stefan Mangard, Paul Kocher, Daniel Genkin, Yuval Yarom, et~al.
\newblock Meltdown: Reading kernel memory from user space.
\newblock {\em Communications of the ACM}, 63(6):46--56, 2020.

\bibitem[LSKA22]{qasmbench}
Ang Li, Samuel Stein, Sriram Krishnamoorthy, and James Ang.
\newblock Qasmbench: A low-level quantum benchmark suite for nisq evaluation and simulation.
\newblock {\em ACM Transactions on Quantum Computing}, 2022.

\bibitem[Mac03]{mackay2003information}
David~JC MacKay.
\newblock {\em Information theory, inference and learning algorithms}.
\newblock Cambridge university press, 2003.

\bibitem[MBZ{\etalchar{+}}12]{mohamed2012improved}
Mohamed Saied~Emam Mohamed, Stanislav Bulygin, Michael Zohner, Annelie Heuser, Michael Walter, and Johannes Buchmann.
\newblock Improved algebraic side-channel attack on aes.
\newblock In {\em 2012 IEEE International Symposium on Hardware-Oriented Security and Trust}, pages 146--151. IEEE, 2012.

\bibitem[{Mic}23]{azure}
{Microsoft Azure}.
\newblock {Azure Quantum}, 2023.

\bibitem[Micnd]{quantumwalks}
Rafaele Miceli.
\newblock {Quantum\_Walks}: Qiskit code to simulate quantum walks on graphs with up to 4 nodes.
\newblock \url{https://github.com/rafmiceli/Quantum_Walks}, [n.d.].

\bibitem[MNW{\etalchar{+}}17]{bench_2017}
Kristel Michielsen, Madita Nocon, Dennis Willsch, Fengping Jin, Thomas Lippert, and Hans De~Raedt.
\newblock Benchmarking gate-based quantum computers.
\newblock {\em Computer Physics Communications}, 220:44--55, 2017.

\bibitem[MOD11]{pulp}
Stuart Mitchell, Michael OSullivan, and Iain Dunning.
\newblock Pulp: a linear programming toolkit for python.
\newblock The University of Auckland, Auckland, New Zealand, 2011.

\bibitem[MOPT12]{synthesis2012ches}
Andrew Moss, Elisabeth Oswald, Dan Page, and Michael Tunstall.
\newblock Compiler assisted masking.
\newblock In {\em International Workshop on Cryptographic Hardware and Embedded Systems}, pages 58--75. Springer, 2012.

\bibitem[MRS{\etalchar{+}}20]{openfermion}
Jarrod~R McClean, Nicholas~C Rubin, Kevin~J Sung, Ian~D Kivlichan, Xavier Bonet-Monroig, Yudong Cao, Chengyu Dai, E~Schuyler Fried, Craig Gidney, Brendan Gimby, et~al.
\newblock Openfermion: the electronic structure package for quantum computers.
\newblock {\em Quantum Science and Technology}, 5(3):034014, 2020.

\bibitem[OBL22]{osan2022quantum}
TM~Os{\'a}n, DG~Bussandri, and PW~Lamberti.
\newblock Quantum metrics based upon classical jensen--shannon divergence.
\newblock {\em Physica A: Statistical Mechanics and its Applications}, 594:127001, 2022.

\bibitem[OC14]{chipwhisperer}
Colin O'Flynn and Zhizhang~(David) Chen.
\newblock Chipwhisperer: An open-source platform for hardware embedded security research.
\newblock Cryptology ePrint Archive, Report 2014/204, 2014.
\newblock \url{https://ia.cr/2014/204}.

\bibitem[Off22]{nqco2022}
The National Quantum~Coordination Office.
\newblock Workshop on cybersecurity of quantum computing, 2022.

\bibitem[OR03]{ofran2003analysing}
Yanay Ofran and Burkhard Rost.
\newblock Analysing six types of protein--protein interfaces.
\newblock {\em Journal of molecular biology}, 325(2):377--387, 2003.

\bibitem[OW12]{oren2012tolerant}
Yossef Oren and Avishai Wool.
\newblock Tolerant algebraic side-channel analysis of $\{$AES$\}$.
\newblock {\em Cryptology ePrint Archive}, 2012.

\bibitem[{Oxf}23]{oqc}
{Oxford Quantum Circuits}.
\newblock {Oxford Quantum Circuits}, 2023.

\bibitem[PDF{\etalchar{+}}21]{pino2021demonstration}
Juan~M Pino, Jennifer~M Dreiling, Caroline Figgatt, John~P Gaebler, Steven~A Moses, MS~Allman, CH~Baldwin, Michael Foss-Feig, D~Hayes, K~Mayer, et~al.
\newblock Demonstration of the trapped-ion quantum ccd computer architecture.
\newblock {\em Nature}, 592(7853):209--213, 2021.

\bibitem[PR13]{prouff2013masking}
Emmanuel Prouff and Matthieu Rivain.
\newblock Masking against side-channel attacks: A formal security proof.
\newblock In {\em Annual International Conference on the Theory and Applications of Cryptographic Techniques}, pages 142--159. Springer, 2013.

\bibitem[{Qis}23]{Qiskit}
{Qiskit contributors}.
\newblock Qiskit: An open-source framework for quantum computing, 2023.

\bibitem[{Rig}23]{rigetti}
{Rigetti}.
\newblock {Rigetti}, 2023.

\bibitem[RS10]{renauld2010algebraic}
Mathieu Renauld and Fran{\c{c}}ois-Xavier Standaert.
\newblock Algebraic side-channel attacks.
\newblock In {\em Information Security and Cryptology: 5th International Conference, Inscrypt 2009, Beijing, China, December 12-15, 2009. Revised Selected Papers 5}, pages 393--410. Springer, 2010.

\bibitem[Sam17]{sampaio96}
Gonçalo Sampaio.
\newblock Code in qasm for quantum circuits and algorithms.
\newblock \url{https://github.com/sampaio96/Quantum-Computing}, 2017.

\bibitem[SHS{\etalchar{+}}14]{cryptoeprint:2014/683}
Ling Song, Lei Hu, Siwei Sun, Zhang Zhang, Danping Shi, and Ronglin Hao.
\newblock Error-tolerant algebraic side-channel attacks using bee.
\newblock Cryptology ePrint Archive, Paper 2014/683, 2014.
\newblock \url{https://eprint.iacr.org/2014/683}.

\bibitem[SJPBL14]{szefer2014cyber}
Jakub Szefer, Pramod Jamkhedkar, Diego Perez-Botero, and Ruby~B. Lee.
\newblock Cyber defenses for physical attacks and insider threats in cloud computing.
\newblock In {\em Proceedings of the 9th ACM Symposium on Information, Computer and Communications Security}, AsiaCCS, June 2014.

\bibitem[SJWK09]{sims2009alignment}
Gregory~E Sims, Se-Ran Jun, Guohong~A Wu, and Sung-Hou Kim.
\newblock Alignment-free genome comparison with feature frequency profiles (ffp) and optimal resolutions.
\newblock {\em Proceedings of the National Academy of Sciences}, 106(8):2677--2682, 2009.

\bibitem[SLM{\etalchar{+}}21]{dnn}
Samuel~A Stein, Ryan L'Abbate, Wenrui Mu, Yue Liu, Betis Baheri, Ying Mao, Guan Qiang, Ang Li, and Bo~Fang.
\newblock A hybrid system for learning classical data in quantum states.
\newblock In {\em 2021 IEEE International Performance, Computing, and Communications Conference (IPCCC)}, pages 1--7. IEEE, 2021.

\bibitem[Soo16]{soos2016cryptominisat}
Mate Soos.
\newblock The cryptominisat 5 set of solvers at sat competition 2016.
\newblock {\em Proceedings of SAT Competition}, page~28, 2016.

\bibitem[STW{\etalchar{+}}22]{stefanazzi2022qick}
Leandro Stefanazzi, Kenneth Treptow, Neal Wilcer, Chris Stoughton, Collin Bradford, Sho Uemura, Silvia Zorzetti, Salvatore Montella, Gustavo Cancelo, Sara Sussman, et~al.
\newblock The qick (quantum instrumentation control kit): Readout and control for qubits and detectors.
\newblock {\em Review of Scientific Instruments}, 93(4), 2022.

\bibitem[SW71]{shanno1971linear}
David~F Shanno and Roman~L Weil.
\newblock “linear” programming with absolute-value functionals.
\newblock {\em Operations Research}, 19(1):120--124, 1971.

\bibitem[Sze18]{szefer2018principles}
Jakub Szefer.
\newblock Principles of secure processor architecture design.
\newblock {\em Synthesis Lectures on Computer Architecture}, 13(3):1--173, 2018.

\bibitem[TS21]{qrng}
Kentaro Tamura and Yutaka Shikano.
\newblock Quantum random numbers generated by a cloud superconducting quantum computer.
\newblock In {\em International Symposium on Mathematics, Quantum Theory, and Cryptography: Proceedings of MQC 2019}, pages 17--37. Springer Singapore, 2021.

\bibitem[WSRW21]{synthesis2021icse}
Jingbo Wang, Chungha Sung, Mukund Raghothaman, and Chao Wang.
\newblock Data-driven synthesis of provably sound side channel analyses.
\newblock In {\em 2021 IEEE/ACM 43rd International Conference on Software Engineering (ICSE)}, pages 810--822. IEEE, 2021.

\bibitem[WSW19]{synthesis2019fse}
Jingbo Wang, Chungha Sung, and Chao Wang.
\newblock Mitigating power side channels during compilation.
\newblock In {\em Proceedings of the 2019 27th ACM Joint Meeting on European Software Engineering Conference and Symposium on the Foundations of Software Engineering}, pages 590--601, 2019.

\bibitem[XES23a]{xu2023classification}
Chuanqi Xu, Ferhat Erata, and Jakub Szefer.
\newblock Classification of quantum computer fault injection attacks.
\newblock {\em arXiv preprint arXiv:2309.05478}, 2023.

\bibitem[XES23b]{xu2023exploration}
Chuanqi Xu, Ferhat Erata, and Jakub Szefer.
\newblock Exploration of power side-channel vulnerabilities in quantum computer controllers.
\newblock In {\em Proceedings of the 2023 ACM SIGSAC Conference on Computer and Communications Security (CCS'23)}, pages 1--15, Copenhagen, Denmark, November 26–30 2023. ACM, New York, NY, USA.
\newblock Available at arXiv: \url{https://arxiv.org/abs/2304.03315}.

\end{thebibliography}
